\documentclass[journal]{IEEEtran}

\usepackage[utf8]{inputenc}
\usepackage{algorithmic}
\usepackage[ruled,vlined,boxed,linesnumbered]{algorithm2e}
\usepackage{amsmath,amsfonts, bm}
\usepackage{graphicx}
\usepackage{array}
\usepackage{textcomp}
\usepackage{stfloats}
\usepackage{url}
\usepackage{verbatim}
\usepackage{xcolor}
\usepackage{multirow}
\usepackage{longtable}
\usepackage{subfigure}
\usepackage{booktabs, makecell, tabularx}
\usepackage{siunitx}
\usepackage{hyperref}

\begin{document}

\title{ToupleGDD: A Fine-Designed Solution of Influence Maximization by Deep Reinforcement Learning}

\author{Tiantian~Chen$^{*}$, 
        Siwen~Yan$^{*}$, 
        Jianxiong~Guo,~\IEEEmembership{Member,~IEEE},
        and Weili~Wu,~\IEEEmembership{Senior Member,~IEEE} 
\thanks{T. Chen, S. Yan, and W. Wu are with the Department of Computer Science, The University of Texas at Dallas, Richardson, TX 75080, USA. (E-mail: tiantian.chen@utdallas.edu; siwen.yan@utdallas.edu; weiliwu@utdallas.edu)}

\thanks{J. Guo is with the Advanced Institute of Natural Sciences, Beijing Normal University, Zhuhai 519087, China, and also with the Guangdong Key Lab of AI and Multi-Modal Data Processing, BNU-HKBU United International College, Zhuhai 519087, China. (E-mail: jianxiongguo@bnu.edu.cn)}

\thanks{\textit{($^{*}$Equal contribution; Corresponding author: Jianxiong Guo.)}}
}


\maketitle

\newtheorem{proposition}{Proposition}
\newtheorem{remark}{Remark}
\newtheorem{claim}{Claim}
\renewcommand{\algorithmicrequire}{\textbf{Input:}}
\renewcommand{\algorithmicensure}{\textbf{Output:}}

\newtheorem{theorem}{Theorem}[section]
\newtheorem{corollary}{Corollary}[theorem]
\newtheorem{lemma}[theorem]{Lemma}
\newtheorem{definition}[theorem]{Definition}

\begin{abstract}
Aiming at selecting a small subset of nodes with maximum influence on networks, the Influence Maximization (IM) problem has been extensively studied. Since it is \#P-hard to compute the influence spread given a seed set, the state-of-the-art methods, including heuristic and approximation algorithms, faced with great difficulties such as theoretical guarantee, time efficiency, generalization, etc. This makes it unable to adapt to large-scale networks and more complex applications. On the other side, with the latest achievements of Deep Reinforcement Learning (DRL) in artificial intelligence and other fields, lots of works have been focused on exploiting DRL to solve combinatorial optimization problems. Inspired by this, we propose a novel end-to-end DRL framework, ToupleGDD, to address the IM problem in this paper, which incorporates three coupled graph neural networks for network embedding and double deep Q-networks for parameters learning. Previous efforts to solve IM problem with DRL trained their models on subgraphs of the whole network, and then tested on the whole graph, which makes the performance of their models unstable among different networks. However, our model is trained on several small randomly generated graphs with a small budget, and tested on completely different networks under various large budgets, which can obtain results very close to IMM and better results than OPIM-C on several datasets, and shows strong generalization ability. Finally, we conduct a large number of experiments on synthetic and realistic datasets, and experimental results prove the effectiveness and superiority of our model.
\end{abstract}


\IEEEpeerreviewmaketitle

\section{Introduction}
\IEEEPARstart{O}{nline} social platforms, such as Twitter and LinkedIn, have shown to be one of the most effective ways for people to communicate and share information with each other. Many companies have turned to social network as a primary way of promoting products, and use “word of mouth” effects to maximize the product influence. To maximize earned pro\mbox{f}its, companies may apply a variety of methods, such as distributing free samples or coupons. Many works have focused on the diffusion phenomenon on social networks.  Kempe \textit{et al.} \cite{Kempe03} first formally def\mbox{i}ned Influence Maximization (IM) problem as a combinatorial optimization (CO) problem, and presented  Independent Cascade (IC) model and Linear Threshold (LT) model to depict the information diffusion process.

It has been proved IM is NP-hard, and the objective  (influence spread) is monotone and submodular under IC and LT models \cite{Kempe03}. Kempe \textit{et al.} \cite{Kempe03} used Greedy algorithm to solve IM, which selects the node with maximum marginal gain of influence spread and can achieve $(1-1/e-\epsilon)$-approximation ratio. However, it is \#P-hard to compute the influence spread of a seed set under both IC \cite{Chen101} and LT model \cite{Chen102}. The hardness of estimating the influence spread lies in the randomness of the probabilistic diﬀusion models, i.e., random choices and diffusion paths. The key to approximate the influence spread is to effectively and efficiently sample diffusion paths. Kempe \textit{et al.} \cite{Kempe03} used Monte Carlo method to simulate diffusion paths, which can obtain good estimations when simulation times are large enough. But it is too time-consuming. Borgs \textit{et al.} \cite{BorgsBCL14} first proposed a novel Reverse Influence Sampling (RIS) technique to reduce the running time. However, RIS still incurs significant computational overheads in practice in order to obtain a good solution. Subsequently, a series of algorithms based on RIS were proposed, such as TIM/TIM+ \cite{Tang14}, IMM \cite{Tang15}, SSA/D-SSA \cite{Nguyen171} and OPIM-C \cite{tang2018online}, which can achieve $(1-1/e-\epsilon)$-approximation solution with high probability when the number of generated random reachable reverse (RR) sets are large enough, and were recognized as the state-of-the-art methods to solve IM. However, these algorithms, such as IMM, still have scalability issues in large inﬂuence networks.

On the other hand, the development of deep learning and reinforcement learning (RL) has blossomed in the last few years, resulting in an increasing number of works addressing CO problem by learning-based methods. A natural question is: can we estimate the influence spread by learnable parametric function and avoid costly sampling random RR sets? The answer is Yes. Khalil \textit{et al.} \cite{KhalilDZDS17} first designed an end-to-end deep reinforcement learning (DRL) framework, S2V-DQN, to solve the common CO problem. Then, Manchanda \textit{et al.} \cite{ManchandaMDMRS20} proposed a supervised deep learning based model for the CO problem, called GCOMB, where IM was used as an example to test the performance. However, the exact value of influence spread is not available, and therefore, no accurate target value can be used for supervised learning. On the contrary, Li \textit{et al.} \cite{Li9769766} presented an end-to-end DRL model, called PIANO, by revising S2V-DQN \cite{KhalilDZDS17}. PIANO is trained on subgraphs of the entire network and then tested on the entire network, which makes it not able to generalize on non-homogeneous networks with different topological characteristics. To address the above drawbacks, we integrate the latest strategies and design a new solution framework for IM.

In this paper, we model the IM problem as a RL problem, which aims to find the optimal policy of selecting $b$ seeds ($b$ action sequences) to maximize the influence spread (cumulative rewards) of these $b$ seeds. However, the exact Q-value in this RL is not available, and therefore deep Q-network \cite{mnih2015human} (DQN) is a natural solution to solve this issue. Instead of using DQN, we use its improvement double DQN (DDQN) \cite{HasseltGS16}, which can avoid the over-optimistic issue of a simple DQN and achieve better performance. On the other hand, except for the network topology structure, the function approximator in DDQN also needs to well capture the crucial influence cascading effects in IM, which makes it more challenging. The cascading effect represents that the activation of a node will trigger its neighbors in a successive manner, forming a diffusion cascade on social networks. This is consistent with the message passing effect in graph neural networks (GNNs) \cite{zhou2020graph}. Therefore, based on these two techniques, in this paper, we propose a novel end-to-end DRL framework, called ToupleGDD (\textbf{T}hree C\textbf{ouple}d \textbf{G}raph Neural Networks with \textbf{D}ouble \textbf{D}eep Q-networks), to solve the IM problem, which incorporates three coupled GNNs for network embedding and DDQN technique for parameter learning. The main contributions can be summarized as follows:
\begin{itemize}
    \item To the best of our knowledge, we are the first to present such an end-to-end framework, ToupleGDD, which combines coupled GNNs and DRL method to effectively solve the IM problem.
    \item We propose a personalized DeepWalk method to learn initial node embedding as input features for the following customized GNN layer, which considers both local and global influence contexts of nodes. 
    \item To capture the crucial cascading effects of information diffusion and network topology, we design three coupled GNNs to learn node embeddings. 
    \item We show that ToupleGDD can be applied on large-scale networks without compromising on solution quality.
    \item Extensive experiments are conducted on synthetic graphs and real-world datasets. Empirical results show that our model can achieve performance very close to IMM, and even outperform OPIM-C on several datasets, which demonstrate the superiority and effectiveness of our proposed model.
\end{itemize}

\textbf{Organization.} Section \ref{RelatedWork} reviews the related works. Section \ref{Preliminaries} presents some preliminaries and framework of the ToupleGDD model. The two main parts of ToupleGDD: network embedding and RL formulation, are introduced in Section \ref{nodeEmd} and Section \ref{RL}, respectively. Section \ref{Experiments} is dedicated to experiments and results. Section \ref{Conclusion} concludes the paper.

\section{Related Works}\label{RelatedWork}
\textbf{IM.} Kempe \textit{et al.} \cite{Kempe03} f\mbox{i}rst formulated IM as a CO problem, and presented a $(1-1/e-\epsilon)$-approximation algorithm, Greedy, by applying Monte Carlo method to estimate expected spread of a seed set. But it is too time-consuming. Borgs \textit{et al.} \cite{BorgsBCL14} made a breakthrough for this issue with the RIS technique, which guaranteed  $(1-1/e-\epsilon)$-approximation solutions and signif\mbox{i}cantly reduced the expected running time. Subsequently, a series of more eff\mbox{i}cient randomized approximation algorithms were proposed, such as TIM/TIM+ \cite{Tang14}, IMM \cite{Tang15}, SSA/D-SSA \cite{Nguyen171}, OPIM-C \cite{tang2018online}, and HIST \cite{guo2020influence}. They not only can provide $(1-1/e-\epsilon)$-approximation solution but also are eff\mbox{i}cient even on billion-size networks, which are state-of-the-art approximation algorithms for IM. Later, these algorithms are widely used to solve variations of IM, such as \cite{yang2022aris} \cite{jin2022im2vec}.

\textbf{ML/RL for CO.}\ Recent advancements of deep learning and RL has resulted in an increasing number of works addressing IM by learning-based methods. Since IM can be formulated as a CO problem, many works aiming for CO problems have used IM as an example to test the performance of their models. Khalil \textit{et al.} \cite{KhalilDZDS17} f\mbox{i}rst proposed a DRL model for CO problems, called S2V-DQN, which utilized the graph embedding method, structure2vec \cite{dai2016discriminative}, to encode nodes states to formulate the value approximator, and the f\mbox{i}tted Q-learning to select the node to add to the current seed set. Li \textit{et al.} \cite{LiCK18} approximated the solution quality by graph convolutional networks, and applied a learning framework based on guided tree search. Manchanda \textit{et al.} \cite{ManchandaMDMRS20} proposed a supervised deep learning based model, GCOMB, for CO problems over large graphs. The key contribution of GCOMB is its hybrid learning model, i.e., combining supervised learning and RL. 
By introducing a supervised learning step into Q-learning framework, GCOMB can predict the quality of nodes and f\mbox{i}lter out "bad nodes" at an early step. Instead of solving CO problems on the entire graph,  \cite{KamarthiVWRT20} and \cite{IrelandM22} are focused on how to prune graph and discover a subgraph which can act as a surrogate to the entire graph. Ireland \textit{et al.}  \cite{IrelandM22} introduced a novel graph pruning algorithm, LeNSE, based on supervised learning and RL. LeNSE learns how to identify a subgraph by removing vertices and edges to signif\mbox{i}cantly reduce the size of the problem, so that heuristics can f\mbox{i}nd a nearly optimal solution of a CO problem with a high likelihood. The f\mbox{i}rst phase of GCOMB can be viewed as the graph pruning, which f\mbox{i}lters out the "bad nodes" and only considers the "good" nodes as the candidate. For readers interested in more works of CO, please refer to \cite{BengioLP21} \cite{ MazyavkinaSIB21} \cite{abs-2008-12248} for detailed reviews.  

\textbf{ML/RL for IM.} Fan \textit{et al.} \cite{FanZSL20} proposed the DRL model for network dismantling problem, FINDER, which aimed to f\mbox{i}nd key players in complex networks, and applied GraphSAGE as the function approximator for DQN. Kamarthi \textit{et al.} \cite{KamarthiVWRT20} utilized deep Q-learning for discovering subgraph and solved the IM problem on the subgrah and utilized the selected influential node set as the seeds on the complete graph. There were some researches \cite{LinLC15} \cite{ AliWC18} \cite{Ali0YC20} focusing on using DRL to solve the competitive IM problem, which aims to f\mbox{i}nd an optimal strategy against competitor to maximize the commutative reward under the competition against other agents. Besides, \cite{YadavNROMT18} \cite{ChenQOAT21} considered the
contingency-aware IM problem, where there is a probability of a node willing to be seed when selected as seed node. Tian \textit{et al.} \cite{TianZMWP19} proposed DIEM model for the topic-aware IM problem, which aims to maximize the activated number of nodes under the specif\mbox{i}c query topics. DIEM modif\mbox{i}ed the structure2vec method \cite{dai2016discriminative} for network embeddings, and utilized DDQN with prioritized experience replay to learn parameters. The work most related to ours is \cite{Li9769766}, which proposed a DRL model, called PIANO, for the IM problem, and presented with small  modif\mbox{i}cation from S2V-DQN \cite{KhalilDZDS17}. 

\textbf{Comparisons of related models to our model.} FINDER model \cite{FanZSL20} was proposed for network dismantling problem, and cannot work on directed graphs and weighted graphs. However, our model can work on undirected graphs and different edge weight settings. GCOMB framework \cite{ManchandaMDMRS20} was based on supervised learning which introduced large extra computational overhead and efforts of hand-crafting the learning pipeline, while our model can learn parameters end-to-end. PIANO method \cite{Li9769766} applied structure2vec to learn node embeddings, while we designed three coupled GNNs to learn the network representation. Additionally, both GCOMB and PIANO are trained on subgraphs of the entire graph, and tested on the rest or the entire network, which makes them graph-specif\mbox{i}c. However, our  ToupleGDD model does not have this limitation and performs well on different training and testing datasets, which shows more generalization ability.

\section{Preliminaries and Framework}\label{Preliminaries}
\subsection{Background}
Social network is usually represented by a directed graph $G=(V, E)$, where $V$ denotes the node (user) set and $E$ is a set of relationships between nodes. For an edge $(u, v)\in E$, $u$ is called the in-neighbor of $v$, and $v$ is called the out-neighbor of $u$. For a node $v$, denote by $N_{in}(v)$ and $N_{out}(v)$ the in-neighbor set and out-neighbor set of $v$, respectively. There are many diffusion models to describe the information propagation process on the social network. Since IC model will be used in our experiments, we will introduce it here.  

\begin{definition}[IC model]
Given $G=(V, E)$ with weight function $p: E\rightarrow [0, 1]$, where $p_{uv}$ represents the propagation probability when $u$ tries to activate $v$ by edge $(u, v)$. IC model considers a timestamped propagation process: (1) Each node has two possible states: active and inactive. (2) Initially, all nodes in seed set $S$ are activated and all other nodes are set inactive. (3) If a node $u$ is f\mbox{i}rst activated at timestamp $t$, then $u$ will try to activate its inactive out-neighbor $v$ at timestamp $t+1$ with successful probability $p_{uv}$. After timestamp $t+1$, $u$ cannot activate any of its out-neighbors. (4) Once a node is activated, it remains active in the following timestamps.
\end{definition}

The diffusion process will continue until there is no more node activated. Given a seed set $S$, denote by $I(S)$ the number of activated nodes when the diffusion process terminates. Let $\sigma(S)$ be the expected number of nodes that can be activated by $S$. That is, $\sigma(S) = \mathbb{E}[I(S)]$ and $\sigma(S)$ is called the influence spread of $S$.

\begin{definition}[Influence Maximization (IM)]
Given a social network $G=(V, E)$, a positive integer $b$ and a diffusion model, IM aims to f\mbox{i}nd a small set $S$ of nodes as seeds with $|S|\leq b$, which has the maximum influence spread.  
\end{definition}

Denote by $\sigma(v; S)=\sigma(S\cup\{v\})-\sigma(S)$ the marginal gain obtained by adding $v$ into a seed set $S$. Let $S_{t}$ be the currently selected seed set. The greedy algorithm will select the node which can achieve the maximum of $\sigma(v; S_{t})$ as the next seed. However, computing the influence spread of a seed set is \#P-hard under the IC \cite{Chen101}, resulting in the diff\mbox{i}culty of calculating the marginal gain. Instead of generating a large number of RR sets like in the state-of-art approximation algorithms, in this paper, we regard IM as an RL problem, which aims to f\mbox{i}nd an optimal policy to select $k$ nodes or $k$ action sequence with the maximum influence spread. In this case, the marginal gain can be considered as the value function in RL, whose value is diff\mbox{i}cult to be obtained in our problem. To address this issue, we approximate the value function (marginal gain) by a parameterized function through DRL method.

\begin{definition}[Learning-based IM Problem]
It can be divided into two phases: \textbf{(1) Learning Phase:} Given a set of training graphs $\mathcal{G}=\{G_{1}, G_{2}, \cdots, G_{c}\}$, diffusion model $\psi$ and influence spread function $\sigma: S\rightarrow \mathbb{R}^{+}$, train a group of parameters $\Theta$ such that $\hat{\sigma}(v, S;\Theta)$ could approximate $\sigma(v;S)$ as accurately as possible. \textbf{(2) Testing Phase:} Given a target social network $G$, the learned parameters $\Theta$ and an integer $b$, solve the IM problem with respect to budget $b$ under some diffusion model $\psi$.

\end{definition}

As a special type of RL, DRL applies deep neural networks for state representation and function approximation for value function, policy, transition model, or reward function. In this paper, we use GNNs to obtain node embeddings and formulate the parameterized function using node embeddings, where all parameters are learned by DDQN.

\subsection{General Framework of GNN}
As an effective framework of nodes embedding learning, GNN usually follows a neighbor-aggregation strategy, where the embedding of a node is updated by recursively aggregating embedding from its neighborhood. Formally, $u$'s embedding at $k+1$-th layer $F_{u}^{(k+1)}$ is updated by:
$$m_{\mathcal{N}(u)}^{(k)}=\text{AGGREGATE}^{(k)}({F_{v}^{(k)}: v\in \mathcal{N}(u)}),$$
$$
F_{u}^{(k+1)}=\text{UPDATE}(F_{u}^{(k)}, m_{\mathcal{N}(u)}^{(k)}),
$$
where \text{AGGREGATE} and \text{UPDATE} are neural networks and $\mathcal{N}(u)$ is $u$'s neighborhood.

\subsection{Framework of ToupleGDD}\label{framework}
\begin{figure*}[htbp]
  \centering
  \includegraphics[width=0.85\linewidth]{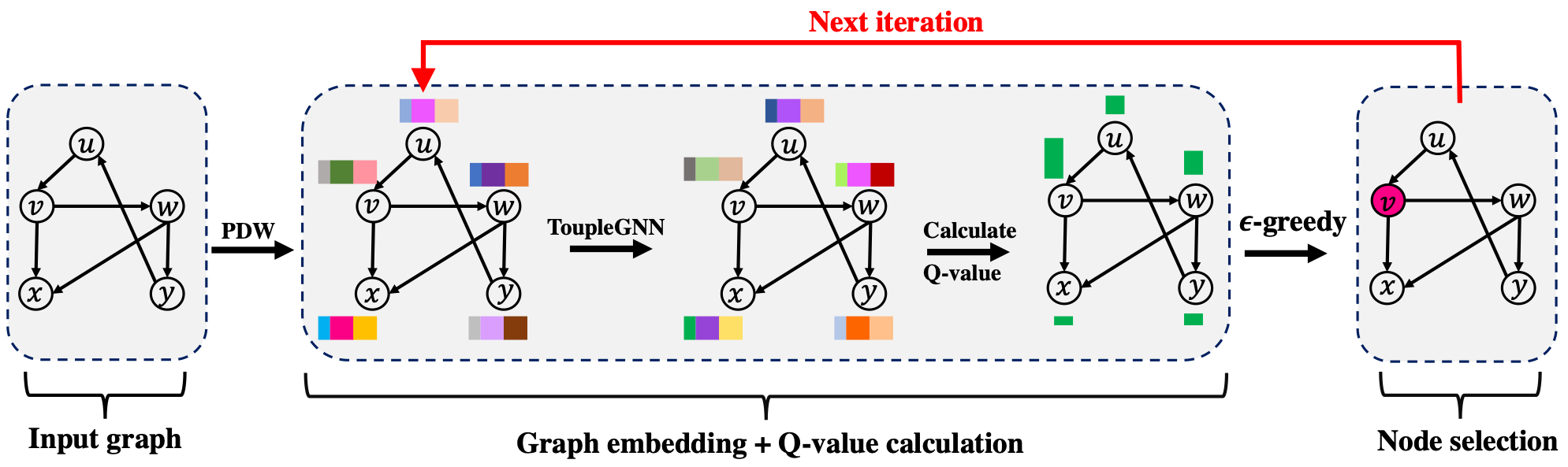} 
  \caption{The framework of ToupleGDD: (a) Apply PDW to obtain initial embedding; (b) Utilize ToupleGNN to capture network topology structures and influence cascading effects to get node embedding; (c) Construct the parameterized function $\hat{Q}(v, S;\Theta)$ based on node embedding input from ToupleGNN; (d) Use $\varepsilon$-greedy to select the next seed and DDQN to learn the parameters.}
  \label{framework2}
\end{figure*}

In this subsection, we present the proposed framework ToupleGDD, which solves the IM problem by incorporating three coupled GNNs and DDQN. The framework of ToupleGDD is illustrated in Fig. \ref{framework2}. Given a set of training graphs $\mathcal{G}=\{G_{1}, G_{2}, \cdots, G_{c}\}$, we f\mbox{i}rst apply the personalized DeepWalk (PDW) method to get the initial node embedding, since it has been found that DeepWalk embedding rather than randomly initialized embedding is vital for stable training of Geometric-DQN, which also works well in our model and will be shown in experiments. Then GNN and attention mechanism are combined to learn node embeddings. Specif\mbox{i}cally, three coupled GNN (ToupleGNN) are designed to capture the cascading effect of information diffusion. After $K$ iterations of ToupleGNN, we use the obtained node embedding to construct the parameterized function $\hat{Q}(v, S;\Theta)$ and use RL technique to learn the parameters. Instead of using DQN, we apply the DDQN to learn the parameters $\Theta$ for $\hat{Q}(v, S;\Theta)$ to approximate the marginal gain $\sigma(v; S)$, and adopt $\varepsilon$-greedy policy to select the next seed. The reason why we use DDQN will also be explained through experiments.

\section{Representation: Node Embedding}\label{nodeEmd}
As a way of representing the node as a vector, node embeddings can capture the network topology. For our IM problem, more importantly, node embeddings need to capture the influence cascading effects, which represents that the activation of a node will trigger its out-neighbors in a successive manner, forming a diffusion cascade on networks. For a target node, whether it will be activated is intrinsically governed by three components: the states of in-neighbors, the influence capacity of in-neighbors and its tendency to be influenced by in-neighbors. In this sense, the cascading effect is intrinsically the iterative interplay between node states, nodes' influence capacity and nodes' tendency to be influenced by others. Therefore, for each node $u$, we include three parts in $u$'s embedding: $X_{u}, S_{u}$ and $T_{u}$, where $X_{u}\in \mathbb{R}$ indicates the activation state of node $u$, $S_{u}\in \mathbb{R}^{l}$ is the capacity of $u$ to influence other users and $T_{u}\in \mathbb{R}^{l}$ is the tendency of being activated by other users.

\subsection{Initial Embedding Learning}
Instead of randomly generating initial embeddings, we proposed the personalized DeepWalk (PDW) method to learn embeddings as input features for the following GNN layer. The main part of PDW is to generate node contexts, and then utilize skip-gram technique to predict contexts for a given node. Inspired by Inf2vec model \cite{FengCKLLC18}, for node $u\in V$, our method includes two parts as $u$'s influence context $C_{u}$: local and global influence context, where local context is a sampled set of nodes that can be activated by $u$ and global contexts are sampled from the $r$-hop out-neighbors of $u$. To limit the size of node contexts, assume length threshold of the node context is $L$ and $\alpha\in [0, 1]$. For a node $u$, we use random walk with restart (RWR) strategy (restart probability is set as 0.15 in this paper) to obtain the local influence context $L_{u}$ of node $u$, and the walk will stop when threshold $\alpha \cdot L$ is reached. After generating local contexts, we randomly sample $(1-\alpha)\cdot L$ nodes from the $r$-hop out-neighbor set $N_{out}^{r}$ of $u$ as global influence context $G_{u}$.

Given a user $u$, the probability of user $v$ being influenced by user $u$, is formulated as a softmax function by their node embeddings: $\Pr(v|u)=e^{X_{u}\cdot S_{u}\cdot T_{v}+X_{v}}/Z(u)$, where $Z(u)=\sum_{w\in V}{e^{X_{u}\cdot S_{u}\cdot T_{w}+X_{w}}}$ is the normalization term. Assume users in $C_{u}$ are independent with each other, then the probability of observing context $C_{u}$ conditioned on $u$'s embedding is $\Pr(C_{u}|u)=\Pi_{v\in C_{u}}\Pr(v|u)$.

We will sample a set of influence contexts, $\mathcal{D}=\{(u_{1}, C_{u_{1}}), \ldots, (u_{q}, C_{u_{q}})\}$ from social network $G$. We consider all the observed influence contexts, and attempt to maximize the log probability of them:
\begin{equation}
\max \sum\nolimits_{(u, C_{u})\in \mathcal{D}}{\sum\nolimits_{v\in C_{u}}{\log \Pr(v|u)}}.
\end{equation}

\noindent
However, it is time-consuming to compute $Z(u)$ directly since we need to enumerate each $w\in V$. In this paper, we utilize the negative sampling technique, which is popularly used to compute softmax functions. Instead of enumerating all nodes, negative sampling method only considers a small set of sampled nodes. For each node $u\in V$, we randomly generate a small set of nodes $N$ as negative instances to approximate the softmax function: 
\begin{equation}\label{approx}
\log \Pr(v|u)\approx \log \sigma(z_{v})+\sum\nolimits_{w\in N}{\log \sigma(-z_{w})},
\end{equation}
where $z_{v}=X_{u}\cdot S_{u}\cdot T_{v}+X_{v}, z_{w}=X_{u}\cdot S_{u}\cdot T_{w}+X_{w}$ and $\sigma(x)=1/(1+\exp(-x))$ is the sigmoid function.

Stochastic Gradient Descent (SGD) method is applied to learn all the parameters. In each step, we update the parameters $\Phi$ by calculating the gradient:
\begin{equation}
\Phi \leftarrow \Phi+\eta \frac{\partial}{\partial \Phi}(\log \Pr(v|u)),
\end{equation}
where $\eta$ is the learning rate and $\frac{\partial}{\partial \Phi}$ represents the gradient of parameters $\Phi$. Based on  (\ref{approx}), the gradient for corresponding parameters can be computed as follows:
\begin{equation}
 \begin{aligned}
  &\frac{\partial}{\partial S_{u}}=(1-\sigma(z_{v}))\cdot X_{u}\cdot T_{v}+\sum\nolimits_{w\in N}{(-\sigma(z_{w}))\cdot X_{u}\cdot T_{w}}\\
  &\frac{\partial}{\partial T_{v}}=(1-\sigma(z_{v}))\cdot X_{u}\cdot S_{u},\ \frac{\partial}{\partial T_{w}}=(-\sigma(z_{w}))\cdot X_{u}\cdot S_{u}\\
  &\frac{\partial}{\partial X_{u}}=(1-\sigma(z_{v}))\cdot S_{u}\cdot T_{v}+\sum\nolimits_{w\in N}{(-\sigma(z_{w}))\cdot S_{u}\cdot T_{w}}\\
  &\frac{\partial}{\partial X_{v}}=1-\sigma(z_{v}),\
  \frac{\partial}{\partial X_{w}}=-\sigma(z_{w})\\
\end{aligned}
\end{equation}

\begin{algorithm}[!t] 
	\caption{PDW}
	\label{PDW}
	Initialize $X_{u}, S_{u}, T_{u}$ by Gaussian distribution $\mathcal{N}(0, 0.01)$\;
    Initialize $W\leftarrow \emptyset$\;
	\ForEach{$u\in V$}{
        $L_{u}\leftarrow \emptyset, G_{u}\leftarrow \emptyset, C_{u}\leftarrow \emptyset$\;
        $L_{u}\leftarrow$ Sample $\alpha L$ nodes by RWR starting from $u$\;
        $G_{u}\leftarrow$ Uniformly sample $(1-\alpha)L$ nodes from $N_{out}^{r}(u)$\;
        $C_{u}\leftarrow L_{u}\cup G_{u}$\;
        Insert $(u, C_{u})$ into $W$\;
	}
    \ForEach{$(u, C_{u})\in W$}{
        \ForEach{$v\in C_{u}$}{
            Update $X_{u}, S_{u}, X_{v}, T_{v}$\;
            Sample a set of negative samples $N$\;
            \ForEach{$w\in N$}{
                 Update $X_{u}, S_{u}, X_{w}, T_{w}$\;        
            }       
        }  
    }
	\Return $ X_{u}, S_{u}, T_{u}$ for each node $u$\;
\end{algorithm}

The proposed PDW method is summarized in Algorithm \ref{PDW}. It contains two parts: influence context generation (lines 3-8) and parameters learning (lines 9-14), which have been illustrated in the above. In the influence context generation part, for each node $u$, local influence context is sampled by RWR strategy, and we use breath first search method to obtain $u$'s $r$-hop out-neighbor set $N_{out}^{r}(u)$ for generating global influence context (upper bounder by $|E|$). Therefore, the time complexity of influence context generation part is $O(|V|(\alpha \cdot L + |E|))=O(|V||E|)$. For the parameters learning part, for each tuple $(u, C_{u})\in W$ (where $|W|=|V|$), $L$ iterations are performed for nodes in $C_{u}$. At each iteration, we f\mbox{i}rst update node embeddings of $u$ and $v$, and then update node embeddings for each node in the negative samples set $N$. Therefore, the running time of the parameters learning part is $O(|V|\cdot L \cdot |N|)=O(|V|)$. Here we consider $L$ and $|N|$ are fixed constants. Thus, the total time complexity of Algorithm \ref{PDW} is $O(|V|+|V||E|)=O(|V||E|)$.

\subsection{ToupleGNN}

\begin{figure*}[htbp]
  \centering
  \includegraphics[width=0.9\linewidth]{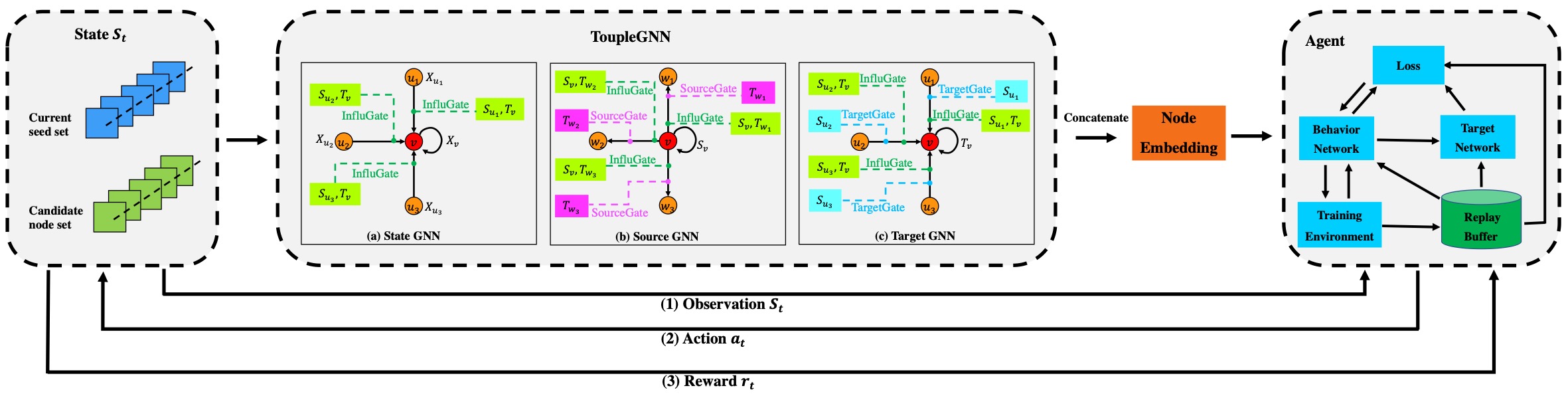} 
  \caption{Mechanism of DDQN incorporated ToupleGNN as function approximator.}
  \label{TripledGNN2}
\end{figure*}

Inspired by \cite{cao2020popularity}, we design three coupled GNNs (ToupleGNN) to naturally capture the iterative interplay between node states, nodes' influence capacity and nodes' tendency to be influenced by others. Taking initial node embeddings as input, ToupleGNN includes three coupled GNNs: (1) state GNN: model the activation states of nodes; (2) source GNN: model the influence capacity of nodes; (3) target GNN: model the tendency of nodes to be influenced by others. The framework of these three GNNs is illustrated in the middle part of Fig. \ref{TripledGNN2}, and we will introduce detailed structures in the following part. Given the currently selected seed set $S_{t}$, we need update node representations accordingly by ToupleGNN.

\subsubsection{State GNN}
The state GNN is used to model the activation state of each node during the cascading effect. For a target user $v$, it will be activated by its active in-neighbors. Therefore, its activation state $X_{v}$ is determined by the activation states of its in-neighbors and the influence weight/probability of these in-neighbors to it. Since the interaction strength between users will change with nodes' states, only using the given static edge weight is not enough to capture the importance and influence weight between users. Therefore,  except for the given edge weights, we also consider applying $v$'s in-neighbors' capacity embedding and $v$'s tendency embedding by an influence attention mechanism to dynamically capture the diffusion weight between them. Specif\mbox{i}cally, def\mbox{i}ne $e_{uv}^{(k)}=\eta^{(k)}[W^{(k)}S_{u}^{(k)},  W^{(k)}T_{v}^{(k)}]$ to measure the dynamic importance of node $u$ to $v$, where $\eta^{(k)}\in \mathbb{R}^{2h^{(k+1)}}$ is a weight vector, $W^{(k)}\in \mathbb{R}^{h^{(k+1)}\times h^{(k)}}$ is a weight matrix to transform the source and target representation from dimension $h^{(k)}$ to $h^{(k+1)}$, and $[\cdot, \cdot]$ denotes the concatenation of vectors. To make coeff\mbox{i}cients comparable among nodes, a softmax function incorporated with the LeakyReLU  \cite{maas2013rectifier} is adopted to normalize the attention coeff\mbox{i}cients:

\begin{small}
\begin{equation}
    \text{InfluGate}(S_{u}^{(k)}, T_{v}^{(k)})=\frac{\exp(\text{LeakyReLU}(e_{uv}^{(k)}))}{\sum_{u\in N_{in}(v)}{\exp(\text{LeakyReLU}(e_{uv}^{(k)}}))},
\end{equation}
\end{small} 

\noindent
where LeakyReLU has negative slope 0.2.

The expected influence that node $v$ aggregates from its in-neighbors is: 

\begin{small}
\begin{equation}
    a_{v}^{(k)}=\sum_{u\in N_{in}(v)}{(\delta_{1}^{(k)}p_{uv}+\delta_{2}^{(k)}\text{InfluGate}(S_{u}^{(k)}, T_{v}^{(k)})})\cdot X_{u}^{(k)}.
\end{equation}
\end{small}

\noindent
Since we expect that the activation state should indicate the possibility of a node being activated, the activation state of node $v$ is set to 1 when it is selected into the current seed set $S_{t}$. Otherwise, $v$'s activation state is updated by aggregating influence from its in-neighbors. That is, node $v$'s activation state at $(k+1)$-th layer is updated by: 
\begin{equation}
    X_{v}^{(k+1)}=
    \begin{cases}
      1, & \text{if}\ v\in S_{t} \\
      \sigma(\xi_{X}^{(k)}X_{v}^{(k)}+\xi_{a}^{(k)}a_{v}^{(k)}), & \text{otherwise}
    \end{cases}
\end{equation}

\noindent
where $\xi_{X}^{(k)}, \xi_{a}^{(k)}\in \mathbb{R}$ are weight parameters and $\sigma(\cdot)$ is the sigmoid function.

\subsubsection{Source GNN}
The source GNN is used to model the capacity of nodes to influence others. Intuitively, the capacity of a node $v$ to activate others can be measured by both its activation state and how much influence its out-neighbors can get when the information is spread from $v$ to out-neighbors, which can be modeled by $v$'s out-neighbors' tendency to be activated.  Similar to the dynamic influence weight def\mbox{i}ned in state GNN, for edge $(v, w)\in E$, we also def\mbox{i}ne the dynamic attention weight $f_{vw}^{(k)}=\beta^{(k)}[W^{(k)}S_{v}^{(k)}, W^{(k)}T_{w}^{(k)}]$ and its corresponding normalization for the weighted aggregation:
\begin{equation}
\alpha_{vw}^{(k)}=\frac{\exp(\text{LeakyReLU}(f_{vw}^{(k)}))}{\sum_{w\in N_{out}(v)}{\exp(\text{LeakyReLU}(f_{vw}^{(k)}}))},
\end{equation}
where $\beta^{(k)}\in \mathbb{R}^{2h^{(k+1)}}$ is a weight vector. Then the neighborhood aggregation is def\mbox{i}ned as:

\begin{small}
\begin{equation}
    b_{v}^{(k)}=\sum_{w\in N_{out}(v)}{(\lambda_{1}^{(k)}p_{vw}+\lambda_{2}^{(k)}\alpha_{vw}^{(k)})\cdot \text{SourceGate}(T_{w}^{(k)})},
\end{equation}
\end{small}

\noindent
where SourceGate($\star$) is the source gating mechanism implemented by a 3-layer MLP in this paper to reflect the nonlinear effect of out-neighbors' target tendency. 

The source representation of node $v$ at $(k+1)$-th layer is updated by incorporating its $k$-th layer source representation, neighborhood aggregation and its activation state:
\begin{equation}
S_{v}^{(k+1)}=\sigma(\gamma_{S}^{(k)}S_{v}^{(k)}+\gamma_{b}^{(k)}b_{v}^{(k)}+\gamma_{X}^{(k)}X_{v}^{(k)}),  
\end{equation}
where $\gamma_{S}^{(k)}, \gamma_{b}^{(k)}, \gamma_{X}^{(k)}\in \mathbb{R}$ are weight parameters.

\subsubsection{Target GNN}
The target GNN is used to model the nodes' tendency to be influenced by others. Generally, the tendency of a node to be activated is determined by its current activation state and the influence diffusion from its in-neighbors to it. Similarly, for edge $(u, v)\in E$, def\mbox{i}ne $d_{uv}^{(k)}=\tau^{(k)}[W^{(k)}S_{u}^{(k)}, W^{(k)}T_{v}^{(k)}]$ and 
\begin{equation}
\phi_{uv}^{(k)}=\frac{\exp(\text{LeakyReLU}(d_{uv}^{(k)}))}{\sum_{u\in N_{in}(v)}{\exp(\text{LeakyReLU}(d_{uv}^{(k)}}))},
\end{equation}
where $\tau^{(k)}\in \mathbb{R}^{2h^{(k+1)}}$ is a weight vector. Then the neighborhood aggregation is def\mbox{i}ned as:

\begin{small}
\begin{equation}
    c_{v}^{(k)}=\sum\nolimits_{u\in N_{in}(v)}{(\rho_{1}^{(k)}p_{uv}+\rho_{2}^{(k)}\phi_{uv}^{(k)})\cdot \text{TargetGate}(S_{u}^{(k)})},
\end{equation}
\end{small}

\noindent
where TargetGate($\star$) is the target gating mechanism implemented by a 3-layer MLP in this paper to reflect the nonlinear effect of in-neighbors' source ability. 

The target representation of node $v$ at $(k+1)$-th layer is updated by incorporating its $k$-th layer target representation, neighborhood aggregation and its activation state:
\begin{equation}
 T_{v}^{(k+1)}=\sigma(\mu_{S}^{(k)}T_{v}^{(k)}+\mu_{c}^{(k)}c_{v}^{(k)}+\mu_{X}^{(k)}X_{v}^{(k)}),  
\end{equation}
where $\mu_{S}^{(k)}, \mu_{c}^{(k)}, \mu_{X}^{(k)}\in \mathbb{R}$ are weight parameters.

\subsection{Putting It Together}
At each iteration of ToupleGNN, information diffusion and network structure features can be passed across nodes. After $K$ iterations, nodes embedding can aggregate information from its $K$-hop neighbors. For node $u$, denote by $X_{u}^{(K)}, S_{u}^{(K)}, T_{u}^{(K)}$ the three components of $u$'s node embedding after $K$ iterations. Then $u$'s node embedding can be obtained by concatenating these three parts: $[X_{u}^{(K)}, S_{u}^{(K)}, T_{u}^{(K)}]$. For the $k$-th layer of state GNN, the time complexity is $O(|V|+|E|)$, which is same for source GNN and target GNN. Therefore, the overall time complexity of ToupleGNN is $O(K(|V|+|E|))$.

Based on the obtained node embeddings, the score function to measure the marginal gain of a node $u\in \bar{S_{t}}=V\setminus S_{t}$ with respect to the current seed set $S_{t}$ is def\mbox{i}ned as $\hat{Q}(u, S_{t}; \Theta)=$

\begin{small}
\begin{equation}\label{q-func}
\theta_{1}^\top\text{ReLU}\biggl(\Bigl[\theta_{2}S_{u}^{(K)}, \theta_{3}\sum_{v\in S_{t}}{S_{v}^{(K)}},\theta_{4}\sum_{w\in V\setminus (S_{t}\cup \{u\})}{T_{w}^{(K)}}\Bigr]\biggr),
\end{equation}
\end{small}

\noindent
where $\theta_{1}\in \mathbb{R}^{3l}, \theta_{2},  \theta_{3}, \theta_{4}\in \mathbb{R}^{l\times l}$ are model parameters. Since the embeddings used to def\mbox{i}ne $\hat{Q}(u, S_{t}; \Theta)$ are computed based on the parameters from ToupleGNN, $\hat{Q}(u, S_{t}; \Theta)$ will depend on $\{\theta_{i}\}_{i=1}^{4}$ and all parameters in ToupleGNN. We will train these parameters (denoted by $\Theta$) end-to-end by RL.

\section{Reinforcement Learning}\label{RL}

\subsection{RL Formulation}
RL concerns about how intelligent agent can take actions according to the current state when interacting with environment to maximize the total reward received. Why do we use RL model to learn the parameters in $\hat{Q}(u, S_{t}; \Theta)$? Actually, IM problem can be naturally formulated as a RL problem: 

\begin{itemize}
    \item Action: an action selects a node $u\in \bar{S_{t}}$ as the next seed, and we use $u$'s node embedding to represent the action.
    \item State: a state $\mathcal{S}_{t}$ represents a sequence of actions of selecting nodes in the current seed set $S_{t}$. We use a $|V|$-dimensional vector to represent state  $\mathcal{S}_{t}$, where the corresponding component of node $u$ is 1 if $u\in S_{t}$, and 0 otherwise. For simplicity, we will use $S_{t}$ instead of $\mathcal{S}_{t}$ to represent the state when there is no ambiguity. The terminal state $S_{b}$ is the state after selecting $b$ nodes.
    \item Transition: changing the activation state $X_{u}$ from 0 to 1 when $u\in \bar{S_{t}}$ is selected as the seed.
    \item Reward: the reward $r(S_{t}, u)$ at state $S_{t}$ is def\mbox{i}ned as the change of reward after selecting node $u$ into the current seed set $S_{t}$ and transition to a new state. That is, $r(S_{t}, u)=\sigma(S_{t}\cup \{u\})-\sigma(S_{t})$ and $r(\emptyset)=0$. In this way, the cumulative reward $R$ of a terminal state $S_{b}$ coincides exactly with the influence spread of seed set $S_{b}$, i.e.,  $R=\sum_{t=0}^{b-1}{r(S_{t}, u_{t})}=\sigma(S_{b})$.
    \item Policy: policy maps a state to possibilities of selecting each possible action. That is, a policy tells the agent how to pick next action. 
\end{itemize}

If we denote by $Q^{*}$ the optimal Q-function for this RL problem, then our embedding parameterized function $\hat{Q}(u, S_{t}; \Theta)$ will be a function approximator for it, which will be learned by DDQN.

\subsection{Training via DDQN}
We use DDQN \cite{HasseltGS16} to perform end-to-end learning of parameters in $\hat{Q}(u_{t}, S_{t}; \Theta)$, which can avoid the over-optimistic issue of a simple DQN by adopting two networks: behavior network and target network, parameterized with $\Theta$ and $\Theta'$, respectively. The target network provides Q-values estimation of future states during training of the behavior network, and only updates parameters $\Theta'$ from the behavior network $\Theta$ every $m$ episodes. The detailed training process is illustrated in Algorithm \ref{Training}. We use the term \textit{episode} to represent a complete sequence of node additions starting from an empty set until termination, and a single action (node addition) within an episode is referred as a \textit{step}. To collect a more accurate estimate of future rewards, $n$-step Q-learning~\cite{sutton1998rlintro} is utilized to update the parameters, which is to wait $n$ steps before updating parameters. Additionally, we apply the fitted Q-iteration~\cite{riedmiller05fittedq} with experience replay for faster learning convergence. Formally, the update is performed by minimizing the following square loss:
\begin{equation}\label{loss}
    (y-\hat{Q}(u_{t}, S_{t}; \Theta))^{2},
\end{equation}
where $y=\sum_{i=0}^{n-1}\gamma^{i} r(S_{t+i}, u_{t+i})+\gamma^{n} \max_{v}{\hat{Q}(v, S_{t+n}; \Theta')}$, and $\gamma\in [0, 1]$ is the discount rate, determining the importance of future rewards.

 Specif\mbox{i}cally, we first apply the PDW method (Alg. \ref{PDW}) to obtain initial embeddings. Then for each episode (Lines 2-20), the seed set is initialized to empty set. For each step, $\varepsilon$-greedy policy is utilized to select a node, which selects a node randomly with probability $\varepsilon$ and with $(1-\varepsilon)$ probability  selects the node with the maximum Q-value (Lines 5-14). If $t\geq n$, it will add the current sample $(S_{t-n}, u_{t-n}, \sum_{i=0}^{n-1}{\gamma}^{i}r(S_{t-n+i}, u_{t-n+i}), S_{t})$ to the replay buffer $M$. Instead of performing a gradient step with respect to the loss of the current example, the parameters are updated with a batch of random samples from the buffer (Lines 24-25). For each episode, we will perform $b$ steps. At each step, node embeddings for each node will be updated for $K$ times by ToupleGNN. At each layer of ToupleGNN, each node aggregates information from its in/out-neighborhood (overall $O(|E|)$). Therefore, the time complexity of each layer is $O(|V|+|E|)$. Putting it all together, the time complexity of Algorithm \ref{Training} is $O(|V||E|+DbK(|V|+|E|))$.

\begin{algorithm}[!t] 
	\caption{Training of ToupleGDD}
	\label{Training}
	Obtain initial embedding for each $u\in V$ by Alg. \ref{PDW}\;
    \For{episode $e$ = 1 to $D$}{
        $S_{0}=\emptyset$\;
        \For{$t$ = 1 to $b$}{
            Uniformly sample a number $c$ from [0, 1)\;
            \If{$c < \varepsilon$}{
                Randomly select a node $u_{t}\in V\setminus S_{t}$\;
            }
            \Else{
                \For{$i$ = 1 to $K$}{
                    \For{$u\in V$}{
                         Update $X_{u}^{(i)}, S_{u}^{(i)}, T_{u}^{(i)}$ by ToupleGNN\;
                    }
                }
                \For{$u\in V$}{
                    Calculate $\hat{Q}(u, S_{t}; \Theta)$ by (\ref{q-func})\;
                }
                Select $u_{t}=\arg \max_{u\in \bar{S}_{t}}{\hat{Q}(u, S_{t}; \Theta)}$\;
            }
            $S_{t}=S_{t-1}\cup \{u_{t}\}$\;
            \If{$t\geq n$}{
               $(S_{t-n}, u_{t-n}, \sum_{i=0}^{n-1}{\gamma}^{i}r(S_{t-n+i}, u_{t-n+i}), S_{t})$ to replay buffer $M$\;
            }       
        }  
        Sample random batch $B\sim M$\; 
        Update $\Theta$ by Adam optimizer over (\ref{loss}) with $B$\;
        Update $\Theta'$ from $\Theta$ every $m$ episodes\;
    }
\Return $\Theta$\;
\end{algorithm}

\section{Experiments}\label{Experiments}
In this section, we conduct several experiments on different datasets to validate the performance of our proposed ToupledGDD model. All experiments are conducted on a machine with Intel Xeon CPU (2.40 GHz, 28 cores), 512 GB of DDR4 RAM, Nvidia Tesla V100 with 16-GB HBM2 memory, running CentOS Linux 7. The source code is available at \url{https://github.com/Dtrycode/ToupleGDD}.

\subsection{Experimental Setup}

\noindent
\textbf{Datasets.} To thoroughly evaluate the performance of the proposed model, both synthetic and real-world datasets are used for evaluation. We generate 20 random Erd\H{o}s-Renyi (ER) graphs with node size varying from 15 to 50 for training and validation. Specif\mbox{i}cally, we f\mbox{i}rst sample the number of nodes uniformly at random from 15 to 50, and then generate an ER graph with edge probability 0.15. Among those generated synthetic graphs, 15 graphs are used for training, and the others are used for validation with the soc-dolphins dataset \cite{Rossi}. The performance of the proposed model and baselines are tested on seven real-world datasets, whose detailed statistics are shown in Table \ref{dataset}. For the undirected graph, we replace each edge with two reversed directed edges. Among these datasets, Twitter, Wiki-1, caGr and Buzznet are from \cite{Rossi}, while Wiki-2, Epinions and Youtube are available on \cite{snapnets}.

\begin{table}[hptb]\centering
\small
\caption{Dataset characteristics}\label{dataset}
\begin{tabular}{p{1.7cm}<{\centering}|p{0.7cm}<{\centering}|p{0.7cm}<{\centering}|p{1.7cm}<{\centering}|p{1.5cm}<{\centering}}
\hline
\textbf{Dataset} & \textbf{\emph{n}} & \textbf{\emph{m}} & \textbf{Type} & \textbf{Average degree}\\
\hline
\emph{soc-dolphins} & 62 & 159 & directed & 5\\
\hline
\emph{Twitter} & 0.8k & 1k & directed & 2\\
\hline
\emph{Wiki-1} & 0.9k & 3k & directed & 6\\
\hline
\emph{caGr} & 4.2k & 13.4k & undirected & 5\\
\hline
\emph{Wiki-2} & 7.1k & 103.7k & directed & 29\\
\hline
\emph{Epinions} & 76k & 509k & directed & 13\\
\hline
\emph{Buzznet} & 101k & 3M & directed & 55\\
\hline
\emph{Youtube} & 1.13M & 3M & undirected & 5\\
\hline
\end{tabular}
\label{tab:dataset}
\end{table}

\noindent
\textbf{Diffusion Models.} Our model can be easily adapted to distinct diffusion models by revising the def\mbox{i}nition of reward function. In this paper, we report the results under the IC model here. Unless otherwise specif\mbox{i}ed,  the probability on edge $(u, v)$ is set to $1/N_{in}(v)$ (in-degree setting), which is widely used in previous works about IM  \cite{Tang14} \cite{Tang15} \cite{Nguyen171}. To fairly evaluate the performance of different methods, we f\mbox{i}rst record the seed set obtained by different methods independently, and then perform 10,000 Monte Carlo simulations to estimate the expected influence spread. All experiments are run 10 times and we report the average of the metric being measured.

\noindent
\textbf{Baselines.}
We compare the performance of ToupleGDD with the state-of-the-art approximation algorithm for IM problem, IMM \cite{Tang15} and OPIM-C \cite{tang2018online}, and the DRL methods S2V-DQN \cite{KhalilDZDS17} and GCOMB \cite{ManchandaMDMRS20} for CO problem. Note that S2V-DQN is originally designed for CO problem, and we revised their code for maximum cut problem to solve IM. Another baseline is PIANO \cite{Li9769766}, which is modif\mbox{i}ed from the S2V-DQN model for IM. For all other baselines, we use the code shared by the authors. For IMM and OPIM-C, we set $\epsilon=0.1$.

\noindent
\textbf{Training and testing details.}
For all training datasets, edge weights are set as in-degree setting. Edge weights on validation datasets and testing datasets have the same setting (we will only specify the setting of testing datasets in the following), and may be set as one of the three settings: (1) in-degree setting; (2) set as 0.1 (0.1-setting); (3) set as 0.5 (0.5-setting). We set the budget $b$ as 5 for all training datasets, while in validation setting 5 and 7 for ER graphs and soc-dolphins dataset, respectively. For each testing dataset, we vary budget $b$ such that $b\in \{10, 20, 30, 40, 50\}$. For S2V-DQN and ToupleGDD, we use RIS method to estimate the influence spread for a given seed set in the training phase. For GCOMB, since their code is not able to deal with multiple training graphs, we follow the same instructions as in their paper and use the training graph shared by them by revising the edge weight to the in-degree setting.

\subsection{Experimental Results}

\subsubsection{Ablation study}
In the early version (called DISCO \cite{abs-1906-07378}) of PIANO model, they have shown that the order of candidate nodes with respect to their Q values remains almost unchanged whenever we select a seed and recompute the network embeddings as well as the Q values. Therefore, instead of iteratively selecting and re-computing nodes embeddings and Q values according to each seed insertion (iterative operation), they simplif\mbox{i}ed the process into only one iteration, by embedding only once and select the top-$b$ nodes with the maximum Q (one-time operation). Inspired by this conclusion and operation, we compare the expected influence spread of seed sets obtained by our ToupleGDD model by these two operations. On the other hand, we also test the impact of the initial embedding to our model. Three groups of experiments are conducted: (1) both train and test with initial embedding (TIEI); (2) train with initial embedding but test without initial embedding (TIEN); (3) both train and test without initial embedding (TNEN). For all of these three types, the validation setting is same as the testing, and all experiments of this part use in-degree probability setting. Besides, for (1) and (2), they share the same training model, and validations are conducted independently for them. For each of the three groups, the iterative and one-time operations are performed at the same one experiment. That is, after computing  the Q-values, we f\mbox{i}rst output the top-$b$ nodes with highest Q-values and then perform the iterative operation according to greedy strategy. Therefore, both of these two operations share the same initial embeddings if there is any. 

The results are shown in Table \ref{initialEmb}. Note that TI and EI represent training and testing with initial embedding, respectively. F\mbox{i}rstly, for same dataset and seeds selection operation (e.g., Twitter with iterative operation), comparing results of three groups, we see that the expected influence spread of seed set obtained by TIEI and TIEN are very close. However, the results of TNEN have big differences from the other two under the same budget, and are not stable under different datasets, which indicates the necessity and importance of initial embedding in training. Secondly, the running time of TIEN and TNEN is less than that of TIEI for same dataset and seeds selection operation, and this difference is significantly big for large datasets, like Epinions. This is because there is no initial embedding generation in TIEN and TNEN when testing, which can save much time especially for large datasets. Besides, the running time of iterative operation increases with the increase of budget due to more iterations and selections, and for one-time operation, there is no significant difference between different budgets. Thirdly, for same group of the experiment (e.g., Twitter under TIEI), comparing the expected spread obtained by iterative and one-time operation, we observe the difference between them is very small, but they actually don't share the exactly same seed set at most cases. However, we cannot f\mbox{i}gure out the reason causing this difference due to the machine accuracy conf\mbox{i}guration for very close values. Besides, one-time operation can output seed set faster than iterative selection, due to its less iteration and computation. From these results, it is convincing that we can use one-time operation for seed selection and TIEN setting to save time but without large decrease of influence spread. For ones who want to apply this algorithm in their problems, it is determined by the trade-off between accuracy and running time.

{\renewcommand{\arraystretch}{0.6}%
\begin{table*}[htbp]\centering
\scriptsize
\caption{Performance of ToupleGDD under different setting }\label{initialEmb}
\begin{tabular}{p{1.0cm}<{\raggedright}|p{1.3cm}<{\raggedright}|p{0.4cm}<{\raggedright}|p{0.4cm}<{\raggedright}|p{2.2cm}<{\raggedright}|p{2.2cm}<{\raggedright}|p{2.2cm}<{\raggedright}|p{2.3cm}<{\raggedright}|p{2.3cm}<{\raggedright}}

\toprule
\multirow{3}{*}{\textbf{Dataset}}&\multirow{3}{*}{\textbf{Operation}}&\multirow{3}{*}{\textbf{TI}} & \multirow{3}{*}{\textbf{EI}}& \textbf{budget: 10} & \textbf{budget: 20} & \textbf{budget: 30}& \textbf{budget: 40} & \textbf{budget: 50}\\
&&&&Influence spread&Influence spread&Influence spread&Influence spread&Influence spread\\
&&&&(time:s)&(time:s)&(time:s)&(time:s)&(time:s)\\
\midrule

\multirow{6}{*}{Twitter}&\multirow{3}{*}{iterative}&$\surd$&$\surd$&147.71 (9.17)&210.86 (16.41)&252.11 (23.42)&287.59 (29.84)&315.88 (38.33)\\
  \cmidrule{3-9}
 &&$\surd$&&146.93 (7.03)&210.66 (13.06)&252.12 (21.13)&287.95 (29.82)&315.99 (33.55)\\
  \cmidrule{3-9}
 &&&&139.36 (5.69)&188.23 (13.80)&234.41 (19.56)&270.09 (27.22)&301.59 (33.03)\\
 \cmidrule{2-9}
 &\multirow{3}{*}{one-time}&$\surd$&$\surd$&146.87 (3.01)&210.86 (3.0)&252.11 (3.08)&288.08 (2.97)&316.60 (3.13)\\
  \cmidrule{3-9}
 &&$\surd$&&147.26 (0.67)&210.66 (0.58)&251.53 (0.64)&287.95 (0.62)&317.28 (0.63)\\
  \cmidrule{3-9}
 &&&&139.36 (0.57)&188.29 (0.66)&232.85 (0.63)&270.11 (0.72)&300.34 (0.73)\\
 
 \midrule
 
  \multirow{6}{*}{caGr}&\multirow{3}{*}{iterative}&$\surd$&$\surd$&213.13 (18.40)&368.74 (25.48)&489.15 (31.60)&602.95 (38.16)&696.95 (46.53)\\
  \cmidrule{3-9}
 &&$\surd$&&214.18 (6.09)&372.13 (12.24)&488.60 (18.15)&602.33 (27.53) &697.99 (37.62)\\
  \cmidrule{3-9}
  &&&&210.56 (5.78)&368.28 (12.89)&489.33 (19.15)&605.18 (27.80)&699.82 (36.81)\\
 \cmidrule{2-9}
 &\multirow{3}{*}{one-time}&$\surd$&$\surd$&208.29 (12.09)&355.87 (11.99)&487.99 (11.65)&604.93 (11.77)&695.95 (11.77)\\
  \cmidrule{3-9}
 &&$\surd$&&210.10 (0.62)&372.79 (0.61)&488.27 (0.57)&608.25 (0.68)&704.79 (0.65)\\
  \cmidrule{3-9}
  &&&&208.24 (0.57)&367.52 (0.63)&487.69 (0.63)&608.23 (0.66)&708.54 (0.67) \\

\midrule

 \multirow{6}{*}{Wiki-2}&\multirow{3}{*}{iterative}&$\surd$&$\surd$&290.48 (46.79)&423.96 (54.68)&521.79 (63.35)&601.39 (71.72)&669.43 (78.23)\\
  \cmidrule{3-9}
 &&$\surd$&&290.81 (7.39)&424.77 (15.79)&523.38 (22.82)&600.29 (31.01)&670.27 (40.04)\\
  \cmidrule{3-9}
 &&&&285.10 (7.24)&422.81 (15.4)&510.53 (19.99)&585.87 (30.18)&647.11 (36.36)\\
 \cmidrule{2-9}
 &\multirow{3}{*}{one-time}&$\surd$&$\surd$&288.97 (39.94)&421.39 (40.46)&518.63 (39.05)&599.93 (39.57)&668.39 (39.16)\\
  \cmidrule{3-9}
 &&$\surd$&&290.22 (0.72)&424.56 (0.70)&516.09 (0.70)&599.04 (0.70)&666.61 (0.73)\\
  \cmidrule{3-9}
 &&&&282.42 (0.67)&420.99 (0.68)&504.32 (0.66)&579.26 (0.72)&642.27 (0.69)\\
 
\midrule

 \multirow{6}{*}{Epinions}&\multirow{3}{*}{iterative}&$\surd$&$\surd$&6022.85 (2.77$\times 10^{3}$)&8303.34 (2.75$\times 10^{3}$)& 9693.69 (2.81$\times 10^{3}$)& 10866.88 (2.79$\times 10^{3}$)&11781.69 (2.8$\times 10^{3}$)\\
  \cmidrule{3-9}
 &&$\surd$&&6018.67 (13.4) &8295.06 (29.9) &9708.72 (44.96) &10853.63 (58.2)&11771.37 (75.61)\\
  \cmidrule{3-9}
 &&&&6013.50 (13.51)&8300.42 (29.11)&9700.72 (42.07)&10840.46 (57.35)&11795.48 (70.3)\\
 \cmidrule{2-9}
 &\multirow{3}{*}{one-time}&$\surd$&$\surd$&6022.85 (2.76$\times 10^{3}$)&8300.36 (2.72$\times 10^{3}$)&9694.49 (2.76$\times 10^{3}$)&10832.49 (2.73$\times 10^{3}$)&11736.67 (2.73$\times 10^{3}$)\\
  \cmidrule{3-9}
 &&$\surd$&&6018.67 (1.36)&8315.56 (1.46)&9718.94 (1.42)&10829.7 (1.4)&11758.36 (1.46)\\
  \cmidrule{3-9}
 &&&&6013.5 (1.38)&8310.07 (1.42)&9729.69 (1.42)&10864.27 (1.43)&11800.19 (1.34)\\
\bottomrule
\end{tabular}
\end{table*}
}

\subsubsection{Influence spread} 
\begin{figure}[!t]
	\centering
	\subfigure[Wiki-2, Performance]{
		\includegraphics[width=0.48\linewidth]{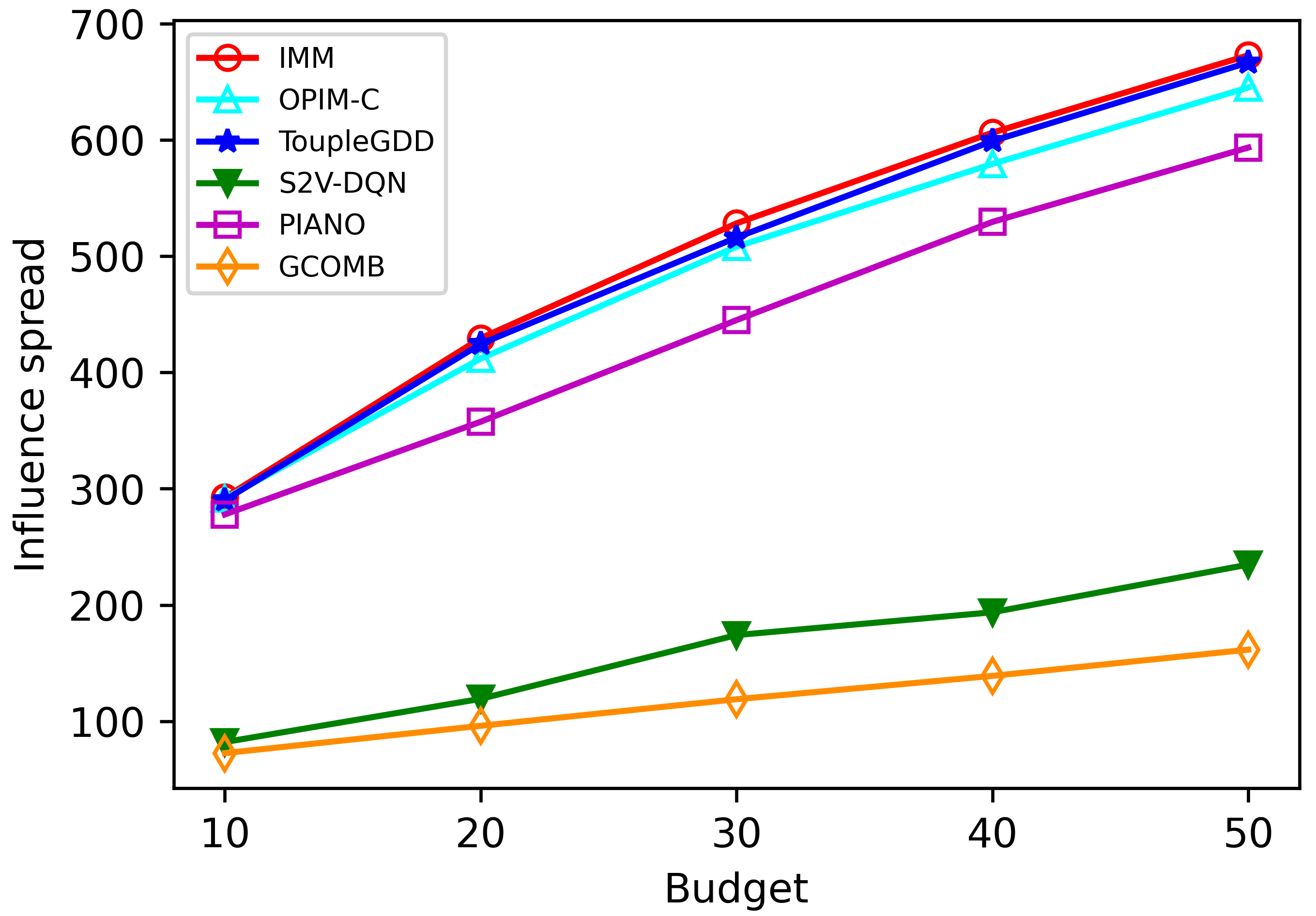}
	}%
	\subfigure[Wiki-2, Running time]{
		\includegraphics[width=0.48\linewidth]{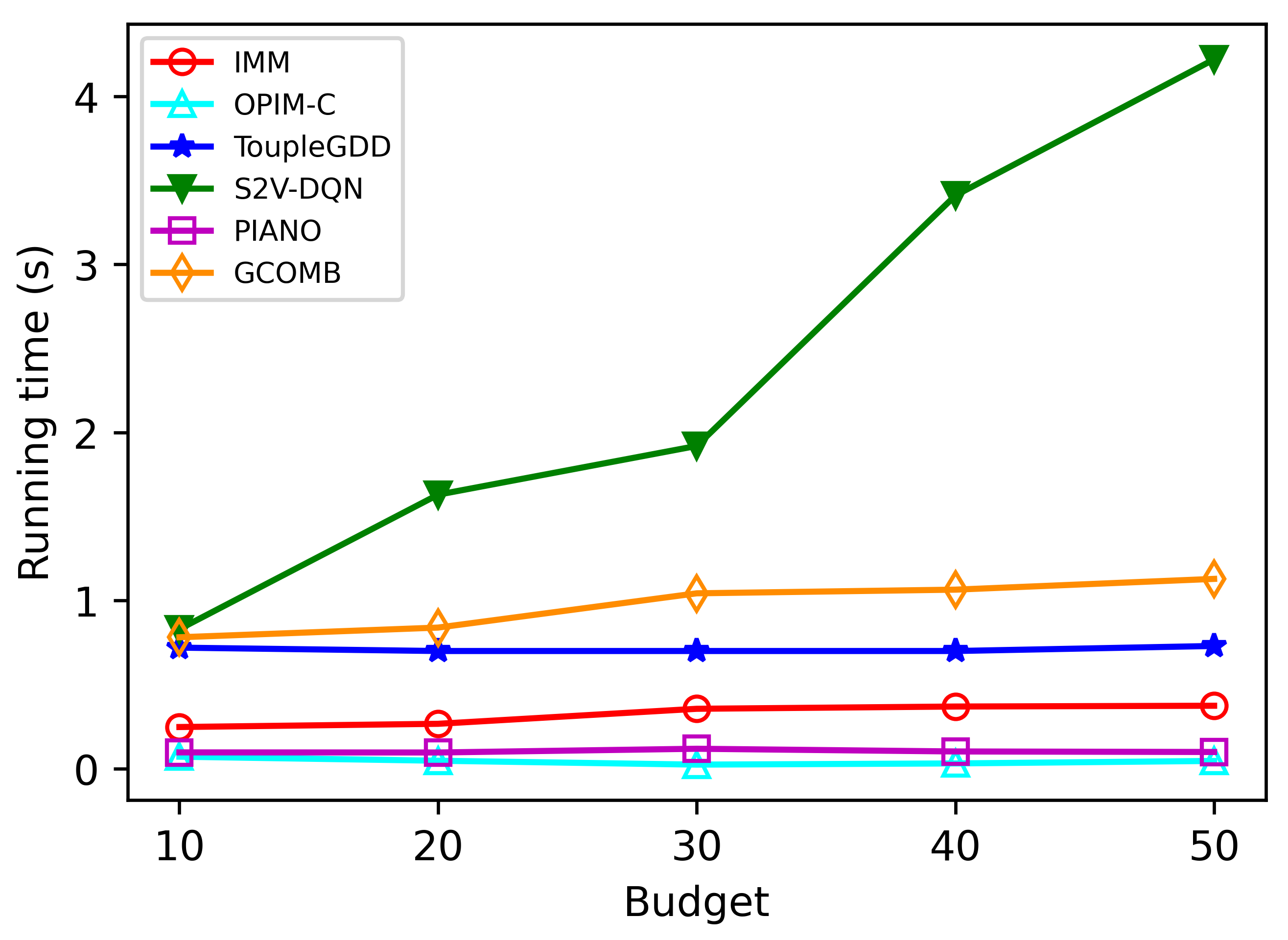}
	}\\
	\centering
	\subfigure[Epinions, Performance]{
		\includegraphics[width=0.48\linewidth]{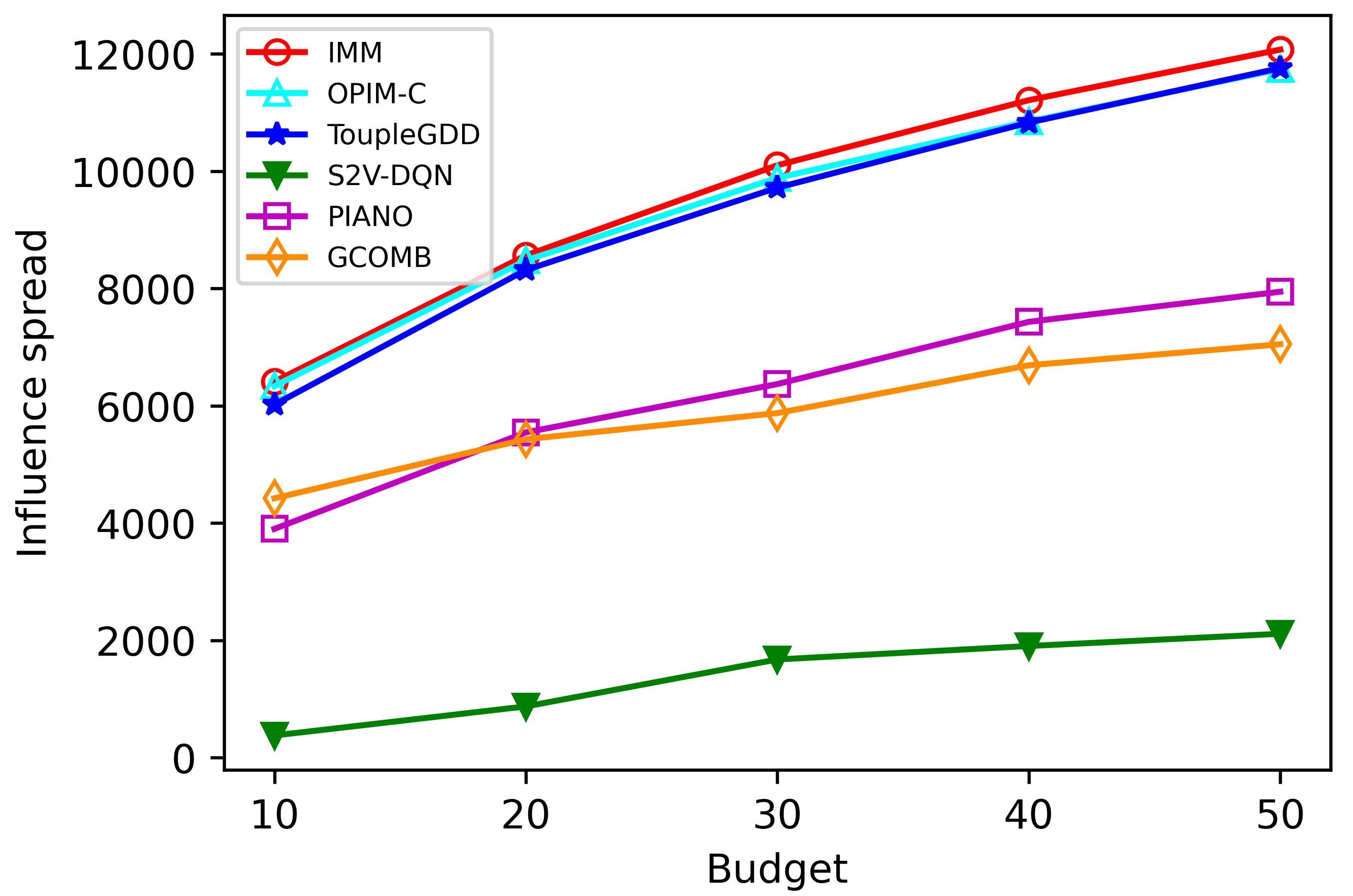}
	}%
	\subfigure[Epinions, Running time]{
		\includegraphics[width=0.48\linewidth]{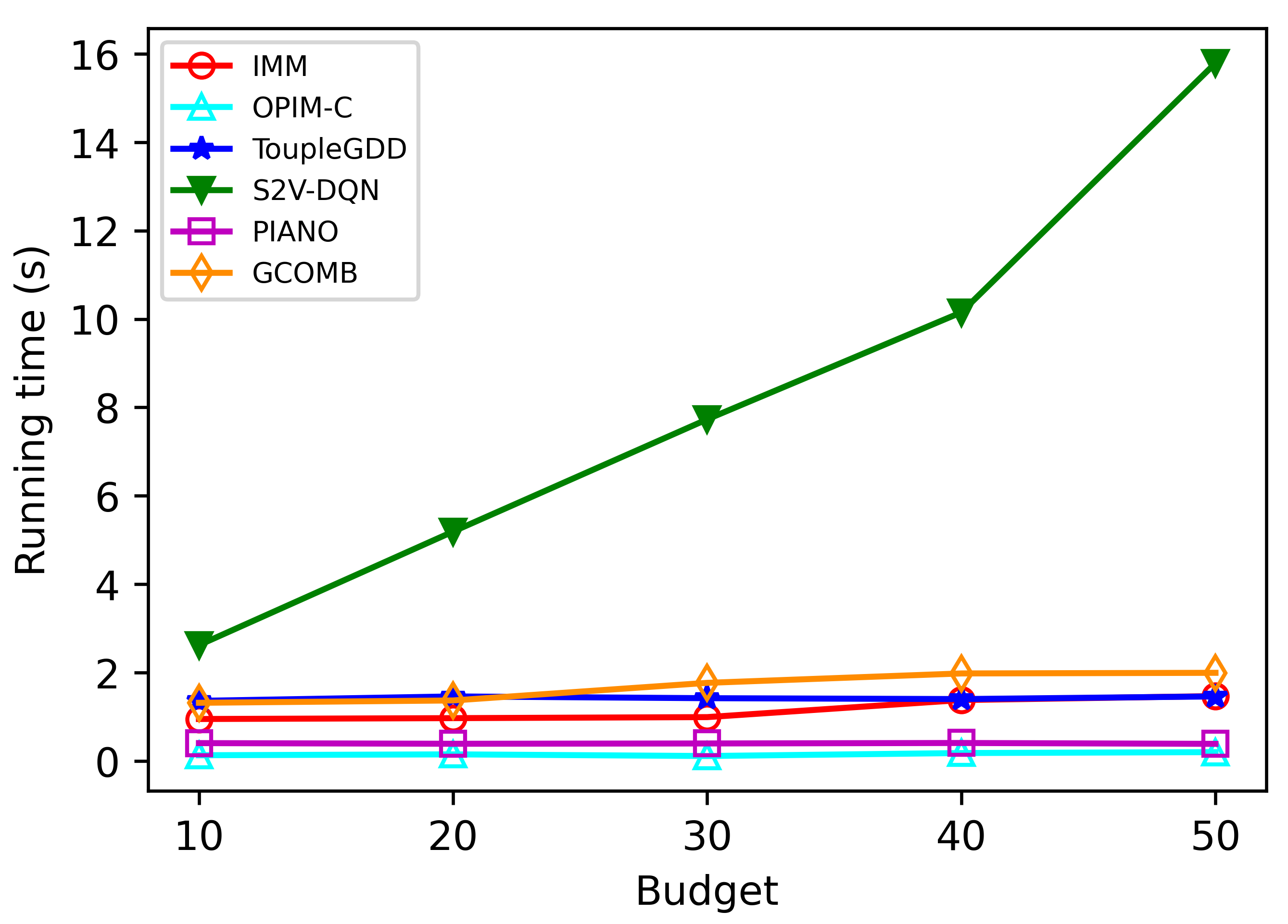}
	}%
	\centering
	
	\subfigure[caGr, Performance]{
		\includegraphics[width=0.48\linewidth]{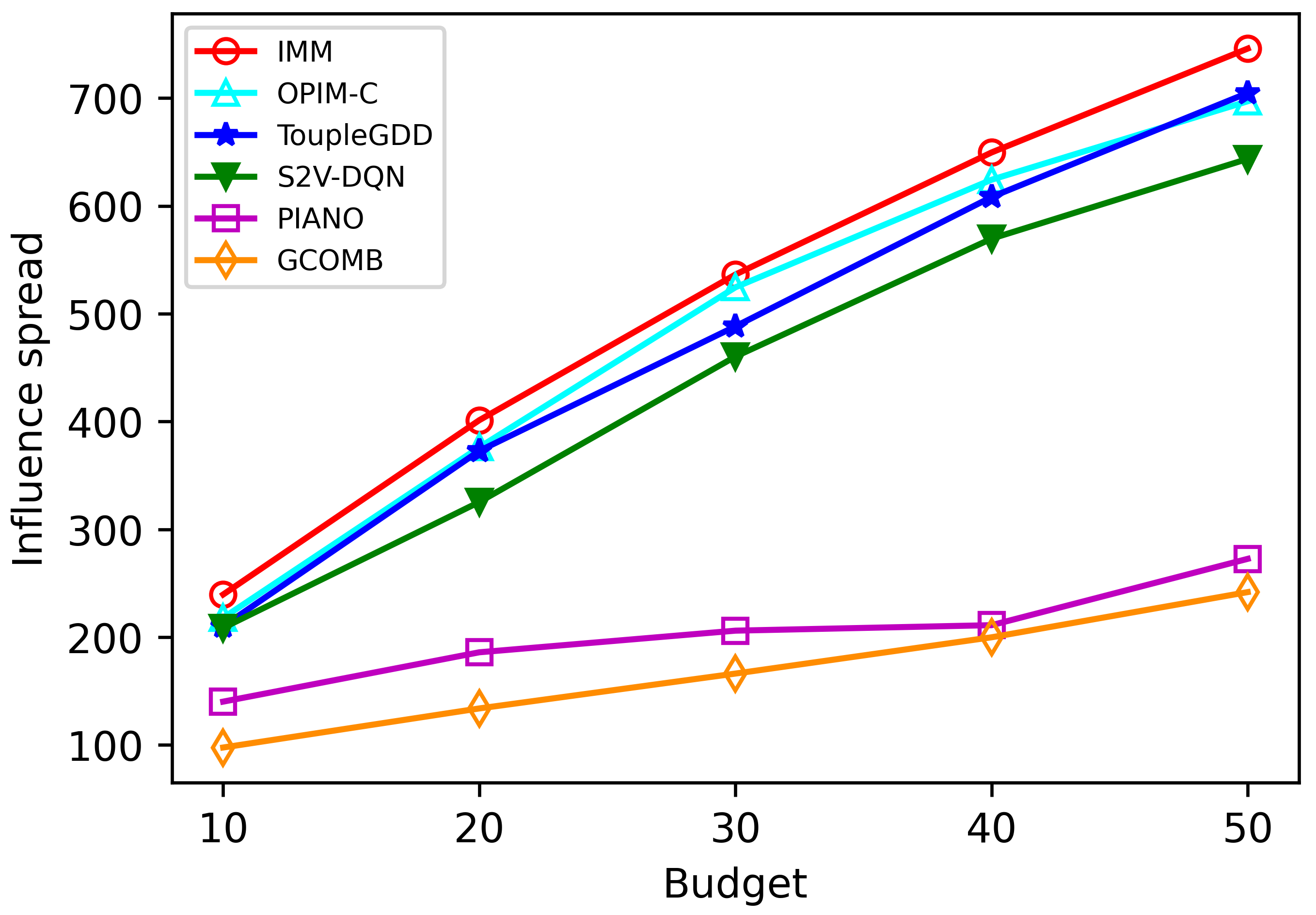}
	}%
	\subfigure[caGr, Running time]{
		\includegraphics[width=0.48\linewidth]{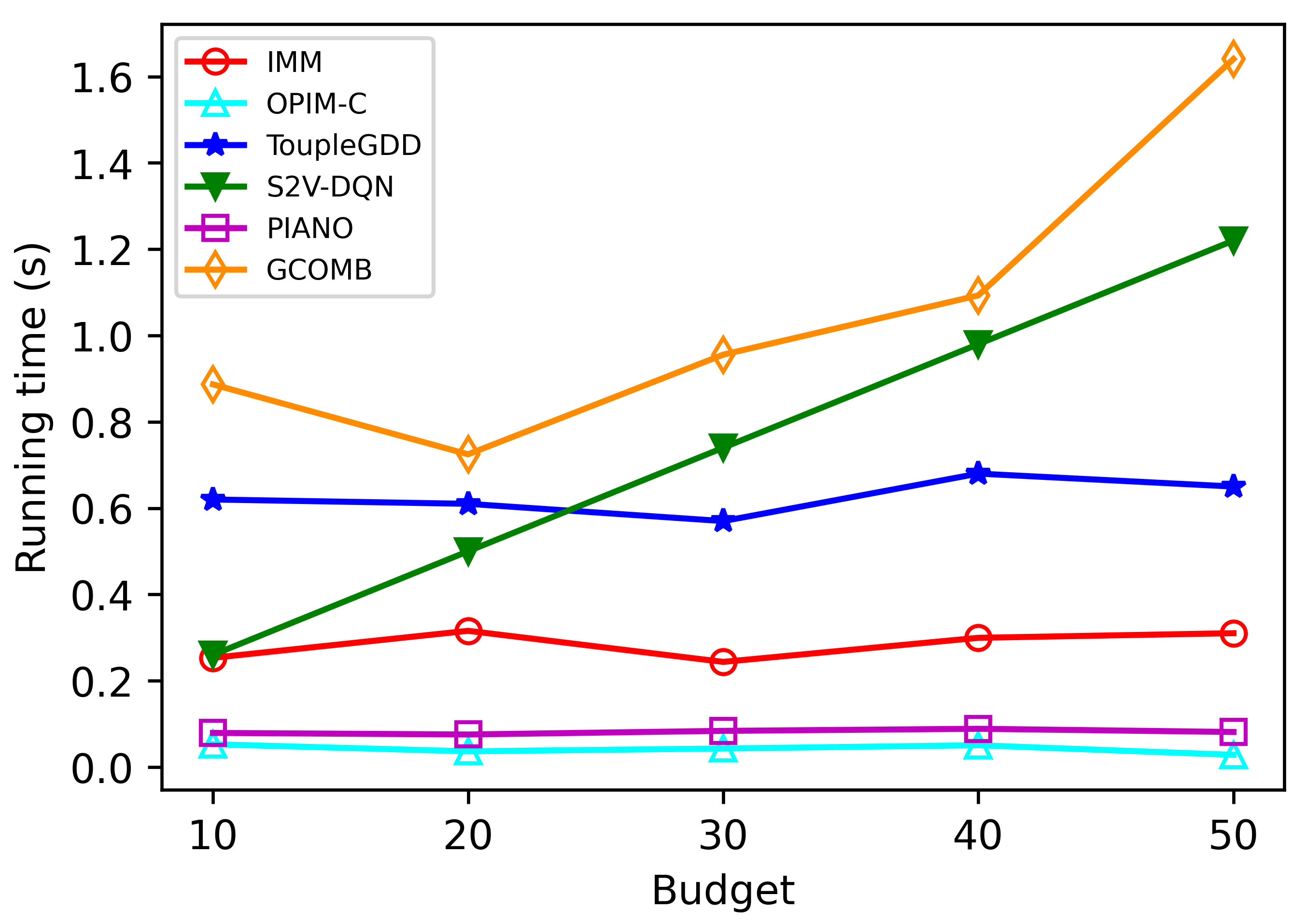}
	}%
 
    \centering
    \subfigure[Buzznet, Performance]{
		\includegraphics[width=0.48\linewidth]{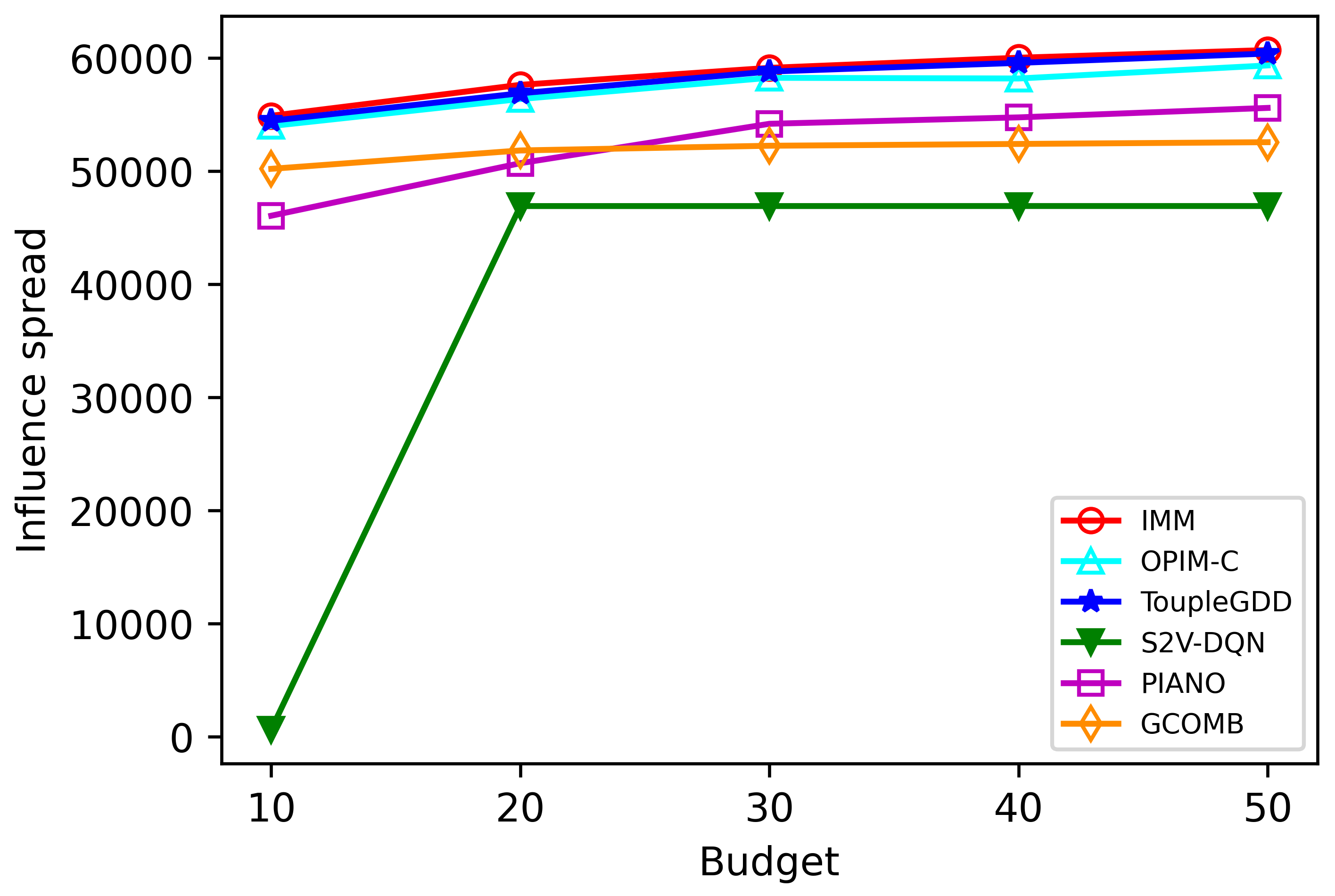}
	}%
	\subfigure[Buzznet, Running time]{
		\includegraphics[width=0.48\linewidth]{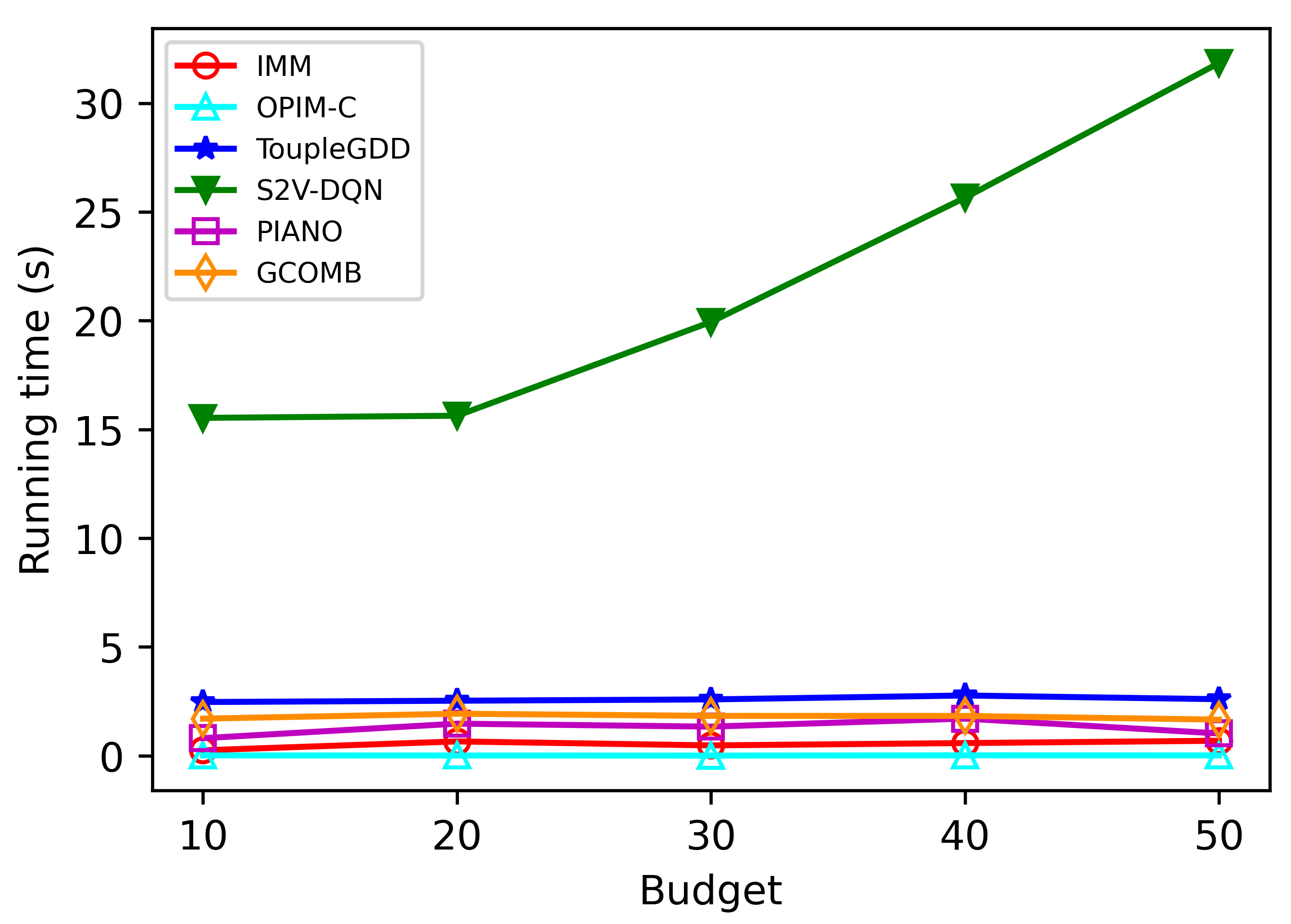}
	}%
	\centering

    \subfigure[Youtube, Performance]{
		\includegraphics[width=0.48\linewidth]{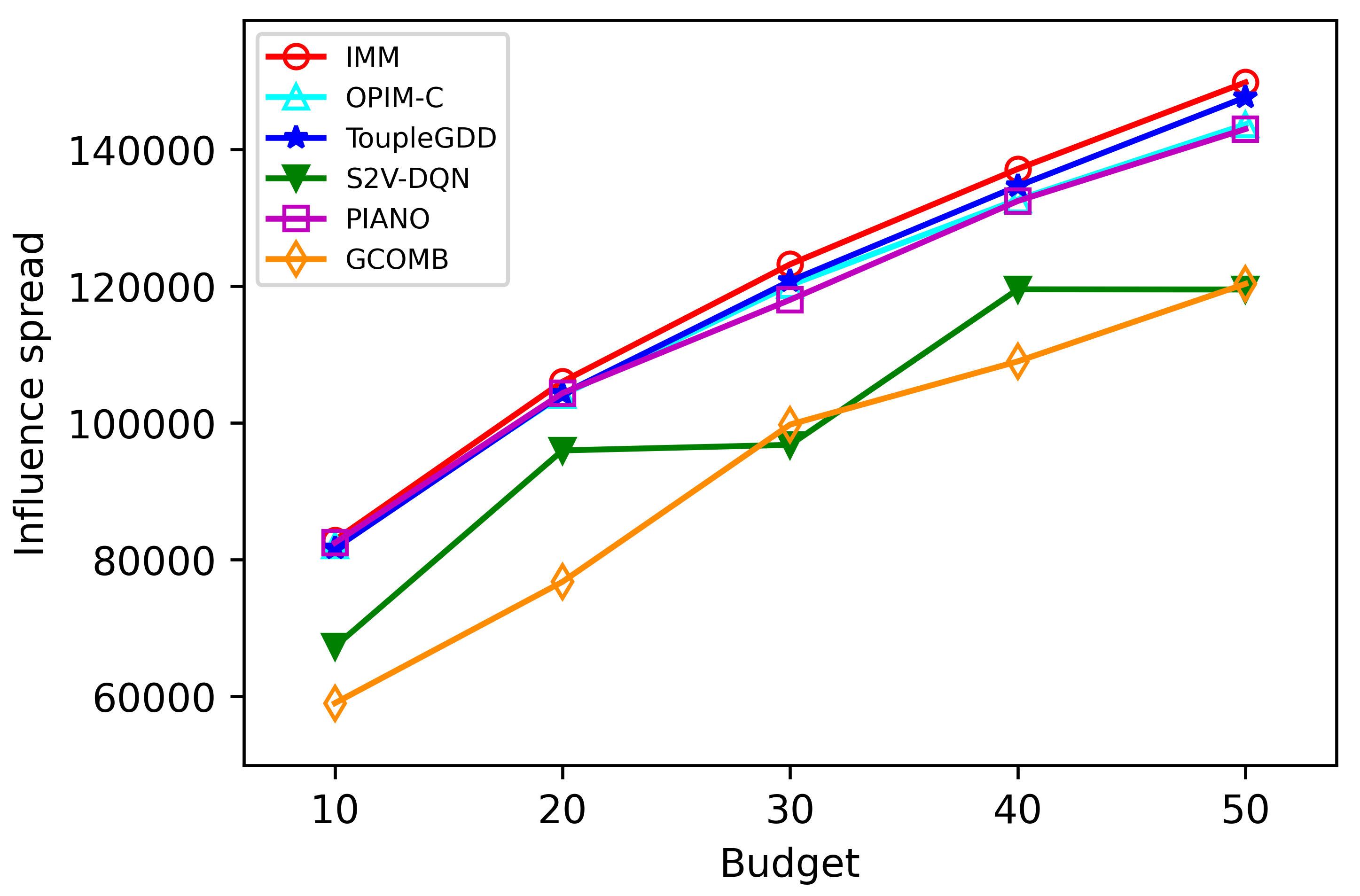}
	}%
	\subfigure[Youtube, Running time]{
		\includegraphics[width=0.48\linewidth]{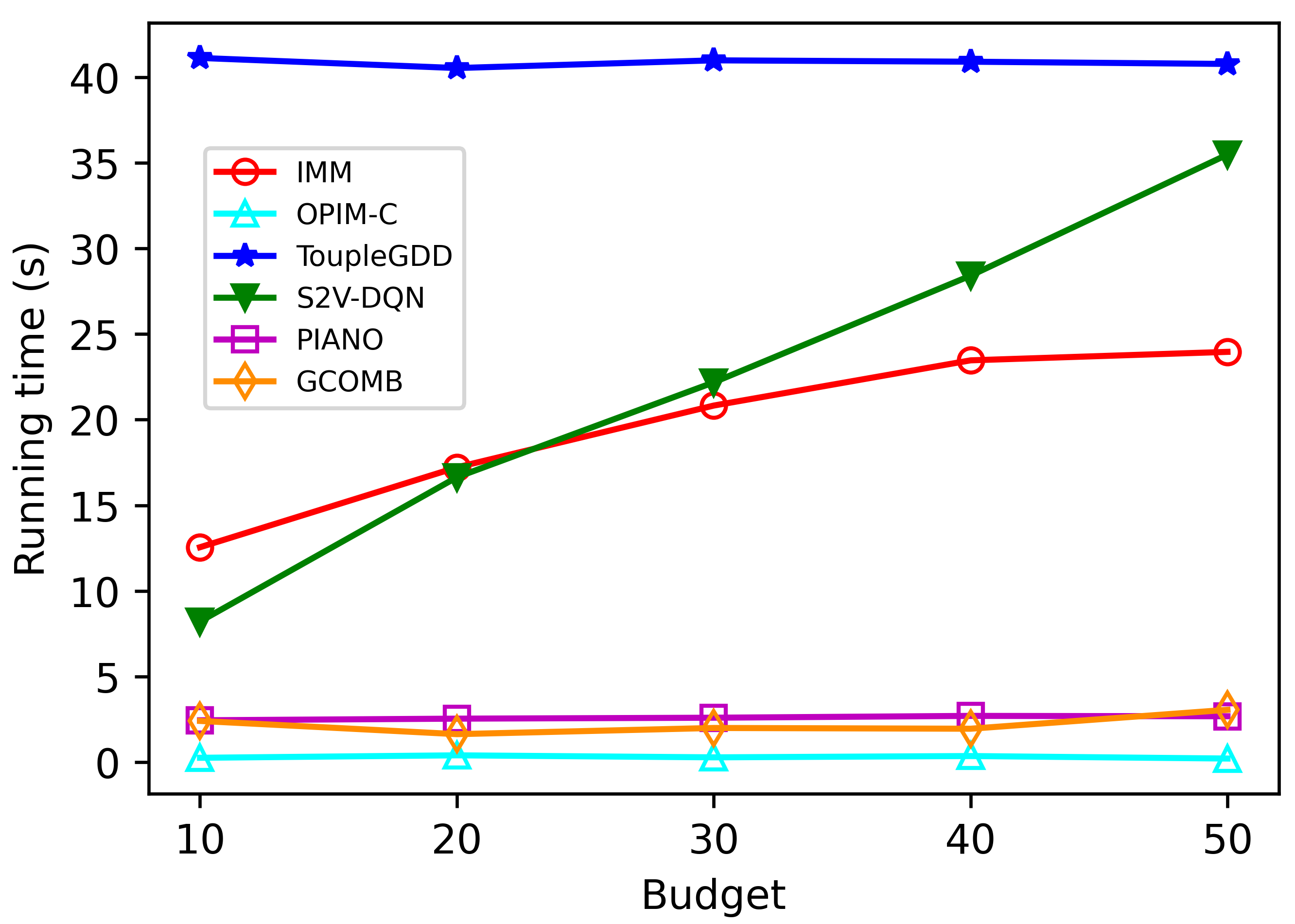}
	}%
    \centering
	\caption{Performance and running time comparisons among different methods.}
	\label{fig3}
\end{figure}

We test the performance of ToupleGDD and baselines on Wiki-1, Epinions, caGr, Buzznet and Youtube datasets with the in-degree probability setting. Fig. \ref{fig3} draws the expected influence spread and running time produced by different models on these five datasets. Note that the results obtained by our model is from TIEN setting and one-time operation, which could not only provide close influence spread with corresponding iterative operation but also runs in less time. From the left column of Fig. \ref{fig3}, the expected influence spread increases with the increase of budget, which is consistent with the monotone increasing characteristic of influence spread under the IC model. Besides, the performance of ToupleGDD is very close to IMM and outperforms OPIM-C on Wiki-2, Buzznet and Youtube datasets, which proves the effectiveness of our model. Comparing the performance of all DRL-based models, ToupleGDD can outperform all other DRL-based models on all tested datasets, demonstrating the superiority of our model. And PIANO and S2V-DQN do not perform stably across different datasets, where S2V-DQN performs better than PIANO on undirected graph caGr, but worse than PIANO on other datasets. This may be because S2V-DQN is designed for undirected graph, and the original paper trained and tested the model on undirected graphs. Even though PIANO is revised from S2V-DQN, its performance is not close to S2V-DQN. This may be because code of PIANO is revised from the code for minimum vertex cover (MVC) problem in the shared S2V-DQN code, while our revised code is from maximum cut (MC) problem. Thus, we use different initial node features. The reason that we choose code of MC is that they have considered edge weight and edge features in MC but not in MVC. Additionally, PIANO and GCOMB have close performance on Epinions, caGr and Buzznet datasets.

\subsubsection{Running time} The right column of Fig. \ref{fig3} draws the corresponding running time of different models to obtain the results in the left column. Note that we only record the time that the model needs to output the seed set for a budget not including the time to compute influence spread of the seed set. We observe that S2V-DQN needs more time to output the seed set than ToupleGDD on all datasets except Youtube. This may be because S2V-DQN use iterative not one-time manner to select seeds, which needs to update embedding and recompute Q-values for $b$ times. Among all the tested methods, OPIM-C needs least time and ToupleGDD model runs a little slower. This may be because our model has many parameters and need to compute the dynamic influence importance between nodes which is time-consuming. But our model's running time is acceptable since it is less than 3 seconds even for million-size dataset Buzznet. Note that this time difference also includes the effects of different implementation language, since ToupleGDD is implemented by Python, while IMM, OPIM-C and most part of PIANO are implemented by C++. We also observe that GCOMB runs slower than ToupleGDD and PIANO on Wiki-2, Epinions and caGr datasets. This may be because GCOMB is proposed for CO problem over very large networks and in their paper, they claimed GCOMB is hundreds of times faster than IMM on million-size datasets. However, from the results in Fig. 3, for small graphs Wiki-2, caGr and Epinions, the running time of GCOMB is longer than ToupleGDD and PIANO due to its extra computational overhead and efforts of hand-crafting the learning pipeline in the supervised learning part.

\subsubsection{Generalization}
\begin{figure}[!t]
	\centering
	\subfigure[Twitter]{
		\includegraphics[width=0.48\linewidth]{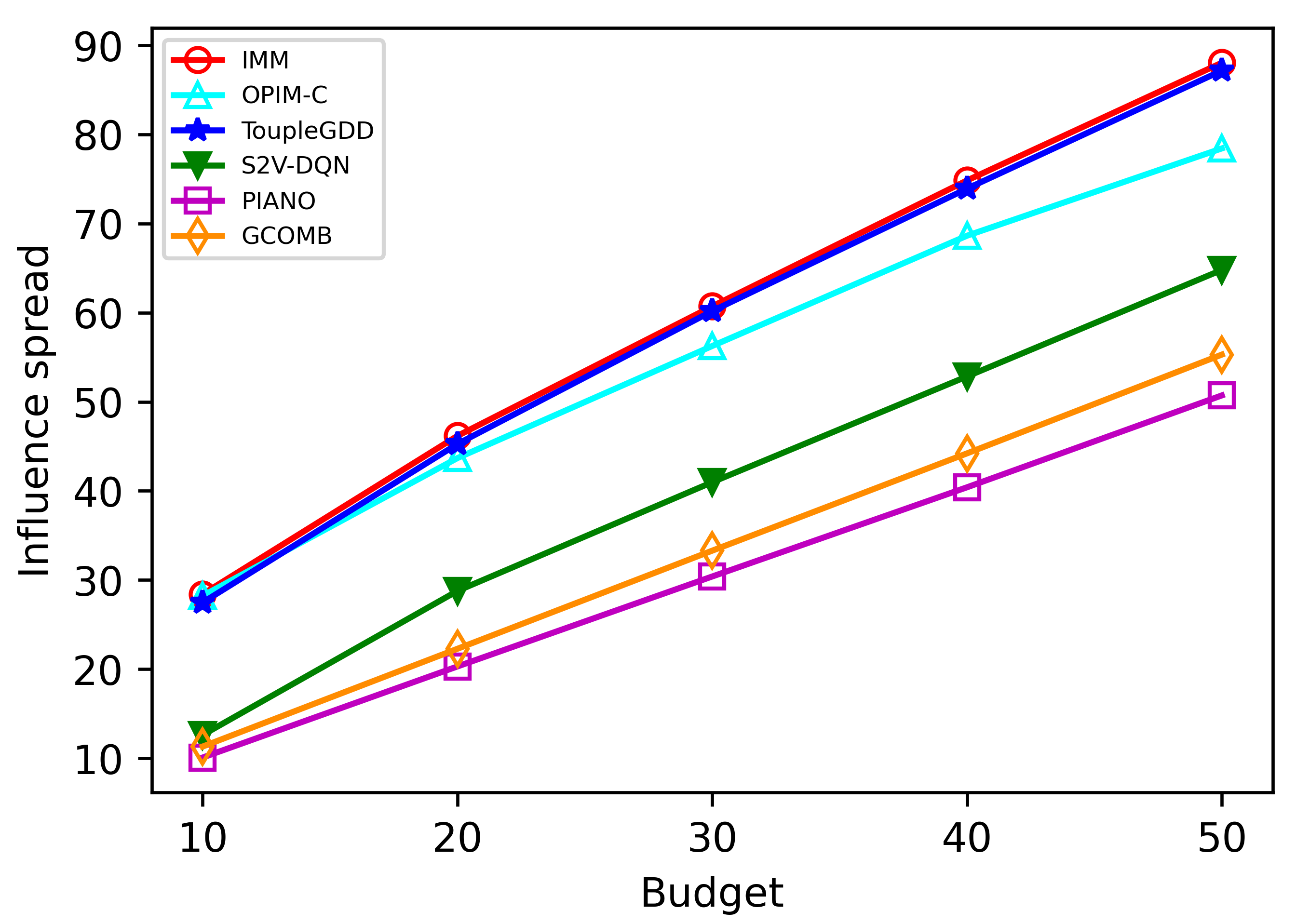}
	}%
	\subfigure[Wiki-1]{
		\includegraphics[width=0.48\linewidth]{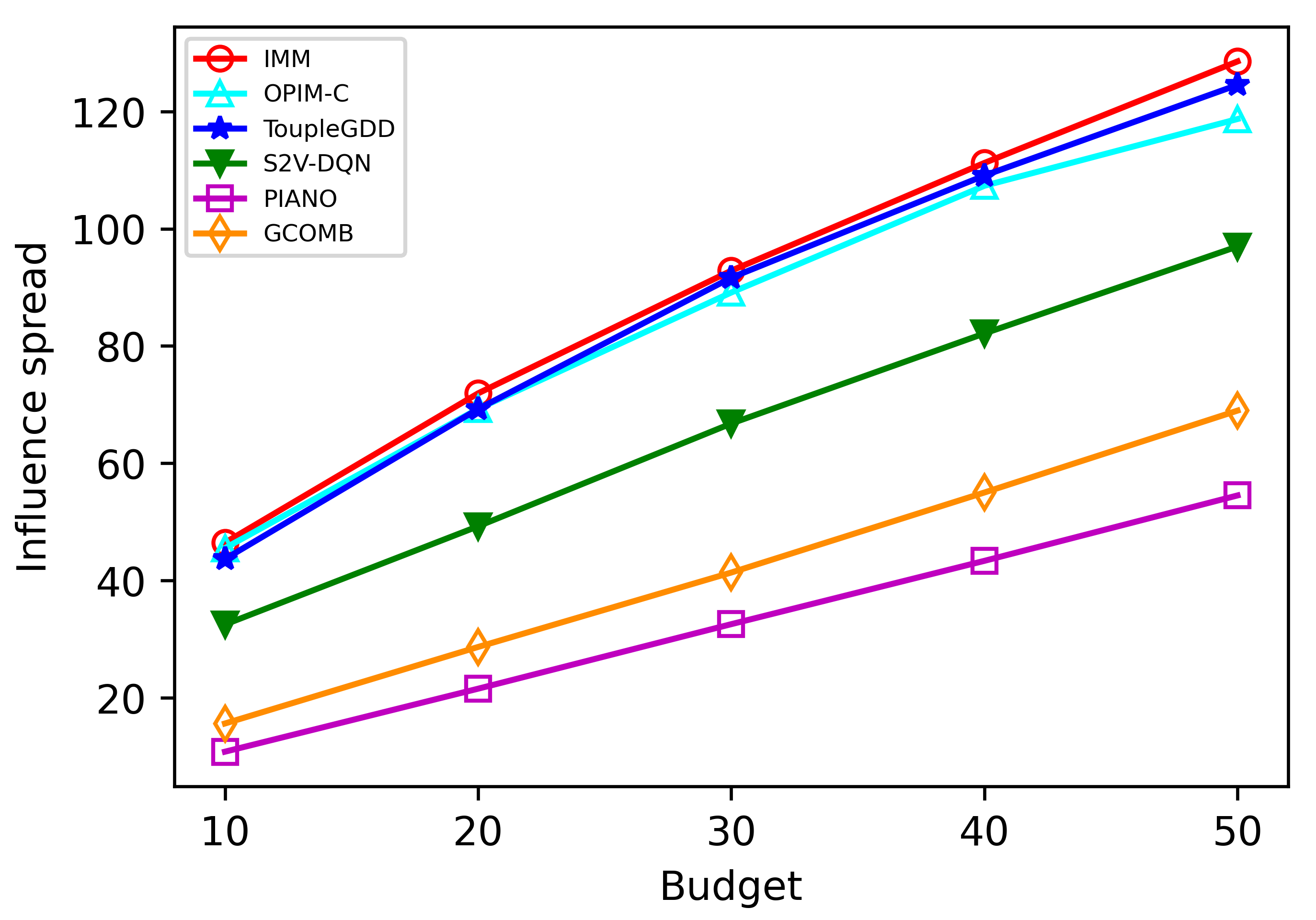}
	}%
	\caption{Performance comparisons among different methods under 0.1-setting.}
	\label{fig4}
\end{figure}

\begin{figure}[!t]
	\centering
	\subfigure[Twitter]{
		\includegraphics[width=0.48\linewidth]{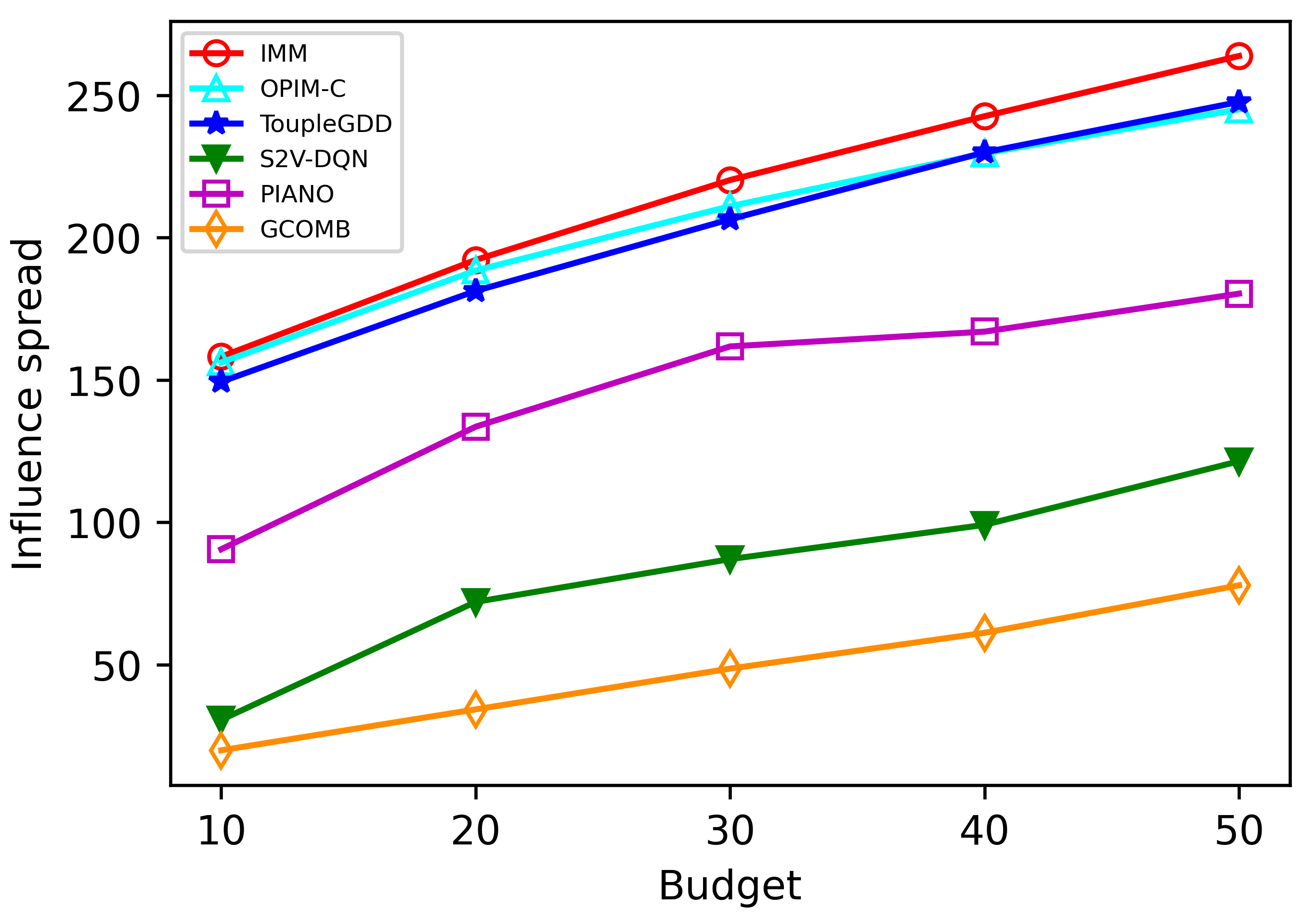}
	}%
	\subfigure[Wiki-1]{
		\includegraphics[width=0.48\linewidth]{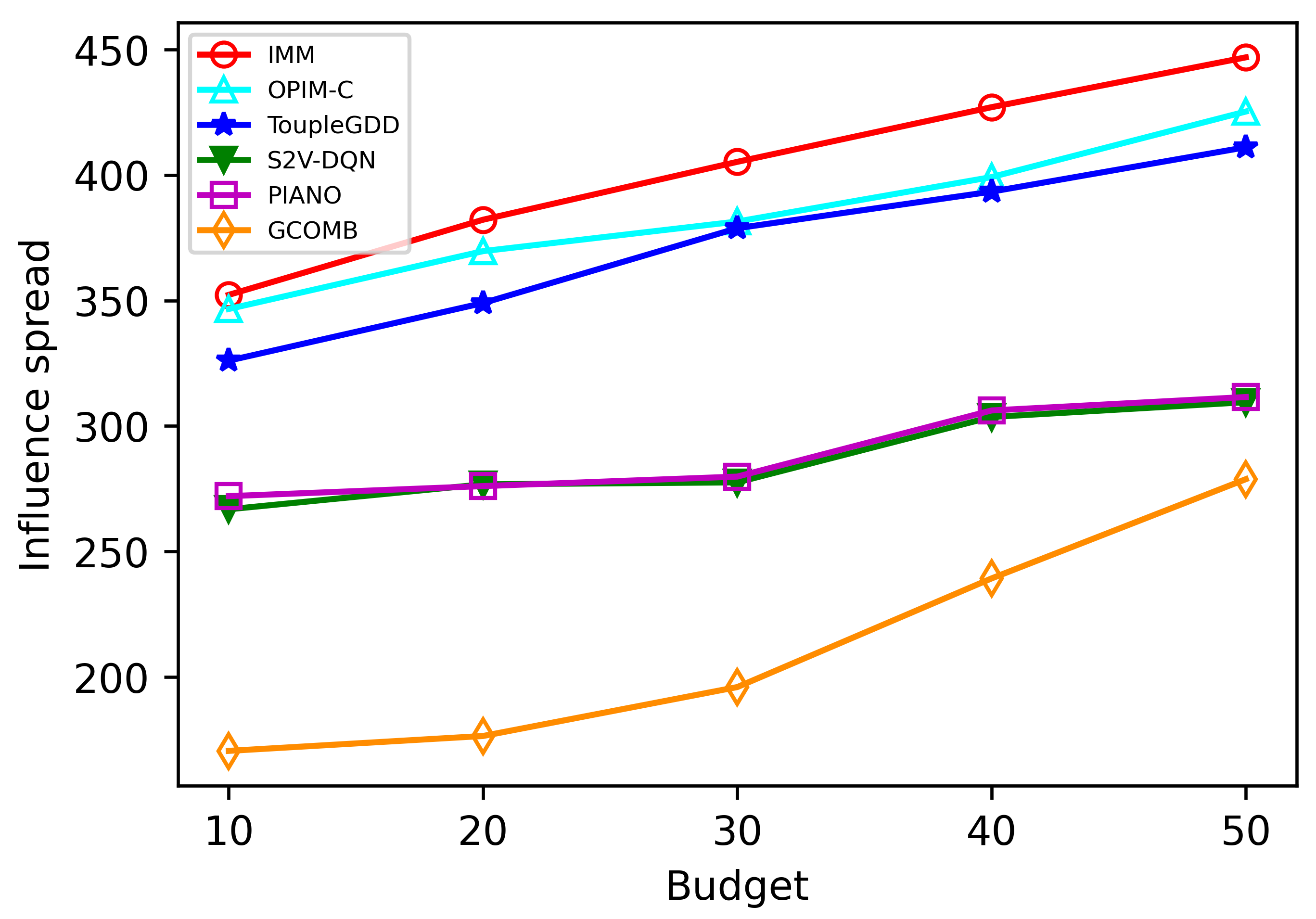}
	}%
	\caption{Performance comparisons among different methods under 0.5-setting.}
	\label{fig5}
\end{figure}

To further validate ToupleGDD's generalization ability, we test the performance of the model trained under in-degree setting and tested on Twitter and Wiki-1 datasets with both 0.1-setting and 0.5-setting. Fig. \ref{fig4} and Fig. \ref{fig5} draw the results for 0.1-setting and 0.5-setting, respectively. From these results, the performance of ToupleGDD is almost equal to IMM and outperforms OPIM-C under 0.1-setting even though it is trained under in-degree setting. And ToupleGDD outperforms all other DRL-based models for both of the two edge probability settings on the two tested datasets. This demonstrates the robustness and generalization ability of the proposed ToupleGDD model. The performance of S2V-DQN, PIANO and GCOMB are not stable across different edge weight settings. S2V-DQN outperforms PIANO and GCOMB under 0.1-setting, while PIANO outperforms S2V-DQN and GCOMB under 0.5 setting. Furthermore, our model can obtain at least 33\% gain of the expected influence spread than S2V-DQN under 0.1-setting, and at least 20\% gain of the expected influence spread than PIANO under 0.5-setting.

\subsection{Intuition of Applying DDQN}
Although DQN is an important milestone for deep learning, several limitations of this algorithm are now known. The improvement to DQN has blossomed in the last decades, such as Dueling DQN (DuelDQN), Double DQN (DDQN) and Duel Double DQN (DuelDDQN). How about the performance of S2V-DQN by replacing DQN with these improvements on IM? This is the main purpose of this group of experiments. We incorporate structure2vec method with these improved models, and compare their performance with S2V-DQN on IM.

We train the four models on soc-dolphins dataset with 0.5-setting for edge weight and the budget is taken from $\{5, 7, 9\}$. Here we keep the budget same for training and testing to avoid the effect of changing budget in the performance. For each budget, we train each framework 1000 epochs with exploration ratio $\varepsilon$ starting from $1$ and multiplied by a factor per epoch to balance exploration and exploitation. We run each framework 5 times to get the average and standard deviation. 

\begin{figure}[!t]
	\centering
	\subfigure[budget 5]{
		\includegraphics[width=0.48\linewidth]{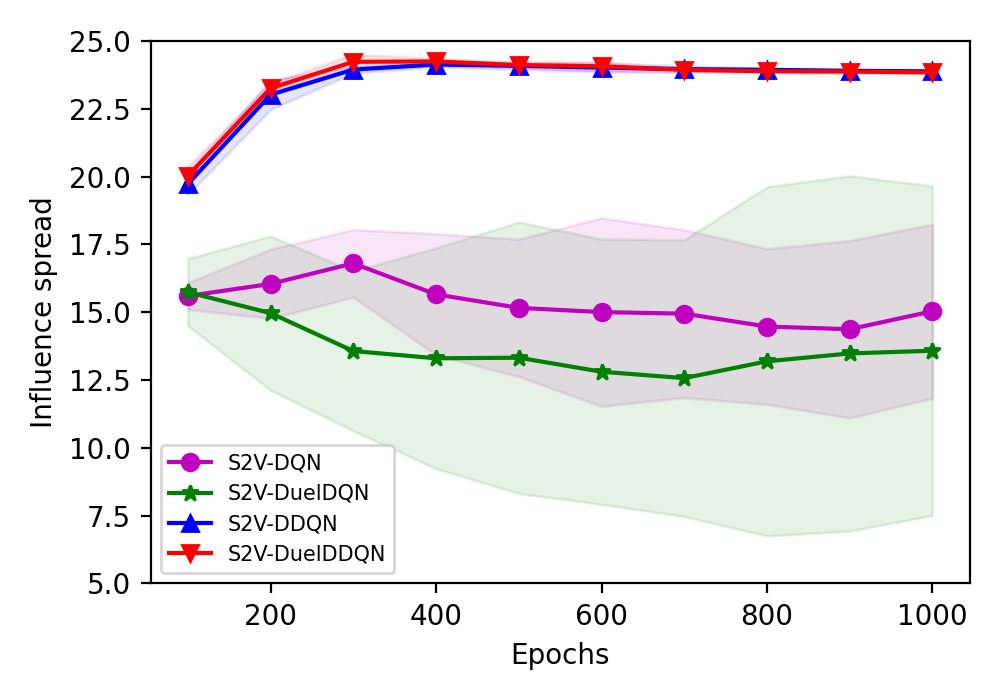}
	}%
	\subfigure[budget 7]{
		\includegraphics[width=0.48\linewidth]{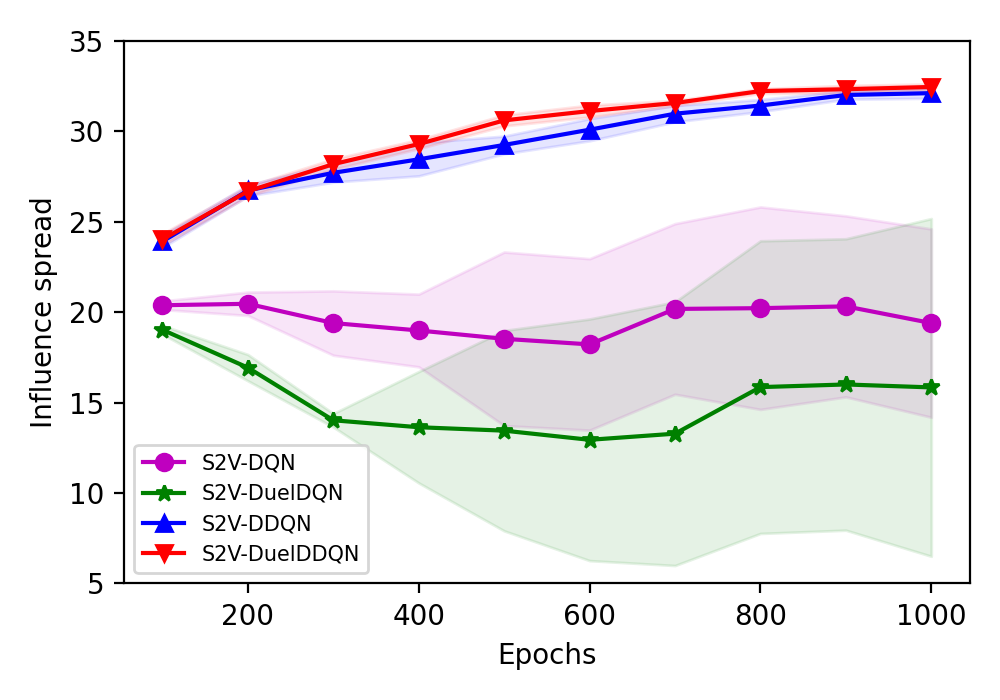}
	}%
	
	\centering
	\subfigure[budget 9]{
		\includegraphics[width=0.48\linewidth]{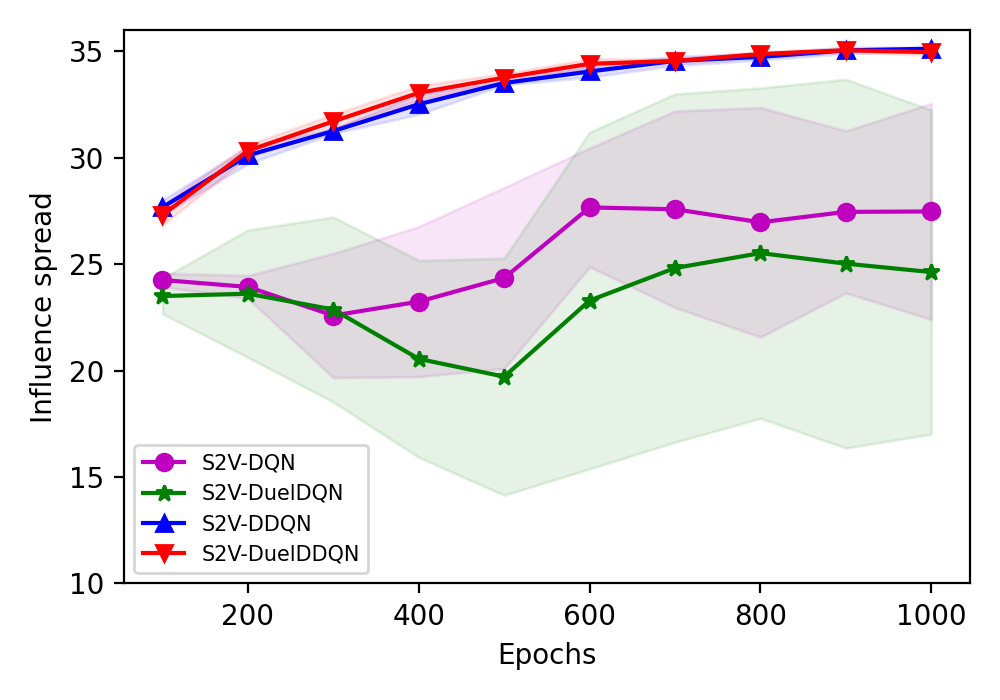}
	}%
	\subfigure[Testing result]{
	\includegraphics[width=0.48\linewidth]{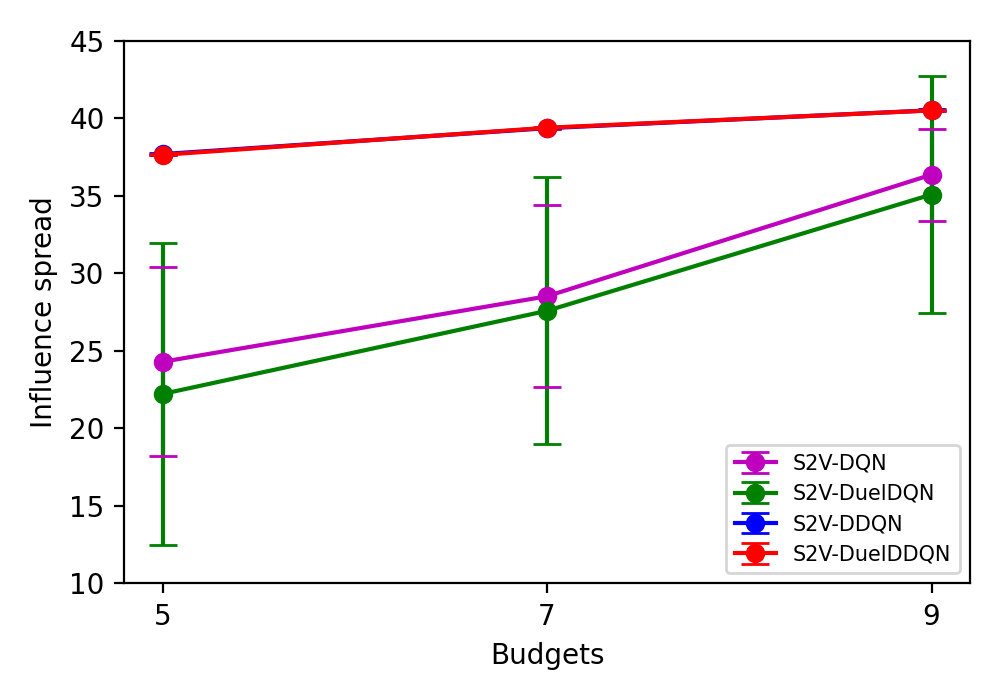}
	}%
	\caption{Training and testing results for different models. (a) (b) and (c): learning curve with budget 5, 7 and 9, respectively; Solid line is average, and shadow is one standard deviation. (d) Testing result: dot is average, and bar shows one standard deviation.}
	\label{learning_curve}
\end{figure}

The learning curves of the four frameworks with budget 5, 7 and 9 are shown in Fig. \ref{learning_curve} (a), (b) and (c), respectively. We expect S2V-DuelDQN to converge fast and S2V-DuelDDQN to perform the best. However, from the results, we observe that S2V-DuelDQN may not work and its advantage of fast converge is not perceivable. The S2V-DDQN does perform well and DuelDDQN manages to make its influence score increase more in fewer epochs. The learning curves fluctuate more with the simple DQN and DuelDQN based models, while the DDQN based models maintain much more stable learning curves across multiple runs.

\begin{table}[thbp]\centering
\scriptsize
\caption{p-value under different budgets (S2V is saved in methods name for space)}\label{p_value_b_5}
\begin{tabular}{c|p{0.6cm}<{\centering}|c|c|c|c}
\toprule
\textbf{Model} & Budget &\textbf{DQN}&{\textbf{DuelDQN}}&{\textbf{DDQN}}&{\textbf{DuelDDQN}} \\
\hline
\multirow{3}{*}{\textbf{DQN}}&5&\multirow{3}{*}{-}& 0.3799& 2.1029e-14 & 2.3740e-14\\
\cmidrule{2-2}
\cmidrule{4-6}
&7&& 0.6611 & 5.7903e-12 & 5.3526e-12\\
\cmidrule{2-2}
\cmidrule{4-6}
&9&& 0.4489 & 1.1655e-08 & 1.2368e-08\\
\midrule
\multirow{3}{*}{\textbf{DuelDQN}}&5&\multirow{3}{*}{-}&\multirow{3}{*}{-}& 4.4682e-10 & 4.8612e-10\\
\cmidrule{2-2}
\cmidrule{5-6}
&7&&& 2.0871e-08 & 1.9727e-08\\
\cmidrule{2-2}
\cmidrule{5-6}
&9&&& 0.0011 & 0.0011\\
\midrule
\multirow{3}{*}{\textbf{DDQN}}&5&\multirow{3}{*}{-}&\multirow{3}{*}{-}& \multirow{3}{*}{-} & 0.0082\\
\cmidrule{2-2}
\cmidrule{6-6}
&7&&& & 0.0457\\
\cmidrule{2-2}
\cmidrule{6-6}
&9&&&& 0.4111\\
\bottomrule
\end{tabular}
\label{tab:p_value_b5}
\end{table}

We test the trained frameworks on a uniformly sampled graph with the same number of nodes and edges as in training dataset. For each model from training, we run 5 times on the testing graph to get its average performance and the seeds are selected with iterative operation. Fig. \ref{learning_curve} (d) draws the expected spread of seed set obtained by different models. We observe that the DDQN  based models still perform better than the simple DQN and DuelDQN model, and are much more stable. Furthermore, we use p-value to check signif\mbox{i}cance of testing performance difference between frameworks as shown in Table~\ref{tab:p_value_b5}. Generally, when p-value is less than 0.05, the performance difference of the two models is significant. The p-value results agree with our previous observation that DDQN based models perform signif\mbox{i}cantly better than DQN and DuelDQN based models. Though the difference between S2V-DDQN and S2V-DuelDDQN is subtle in Fig.~\ref{learning_curve} (d), DuelDDQN does get significantly better performance when the budget is small.

\begin{figure}[!t]
	\centering
	\subfigure[Training time]{
		\includegraphics[width=0.48\linewidth]{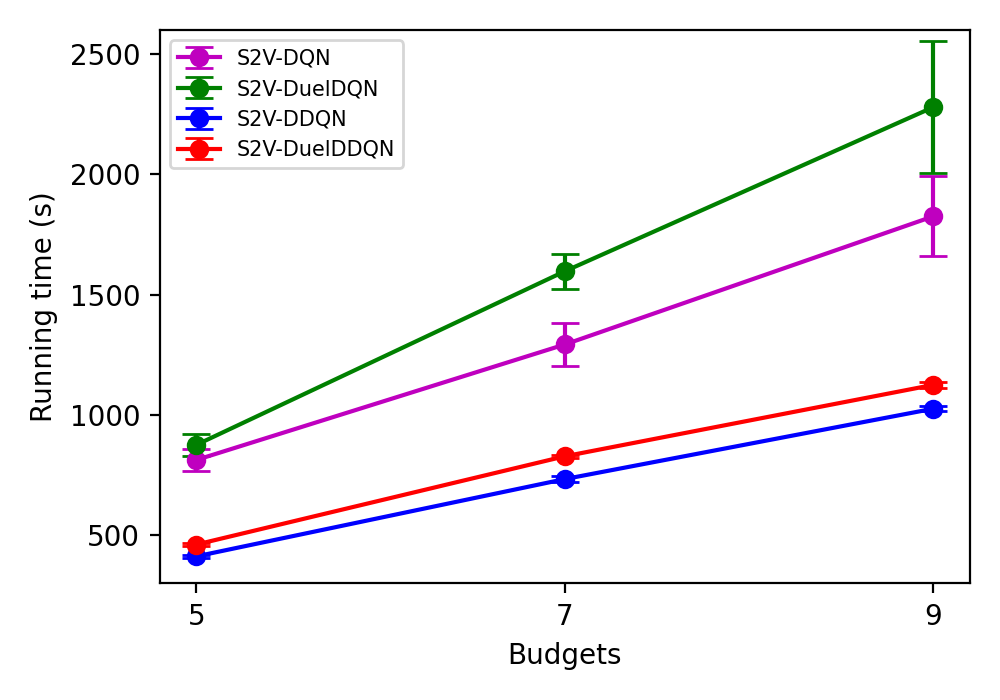}
	}%
	\subfigure[Testing time]{
		\includegraphics[width=0.48\linewidth]{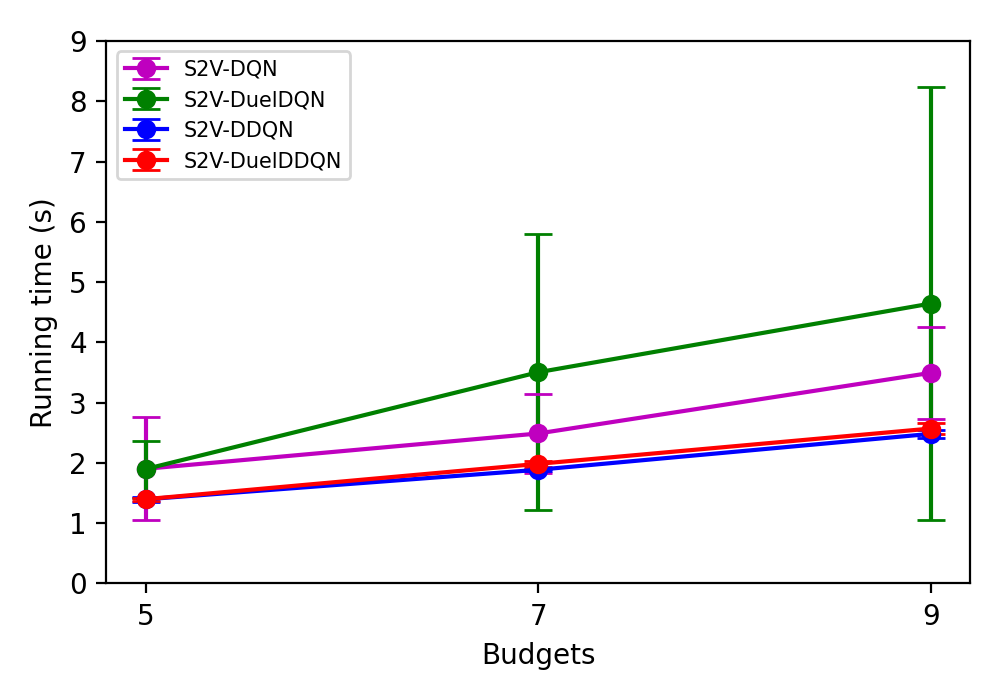}
	}%
	\caption{Running time: dot is average, and bar shows one standard deviation.}
	\label{running_time}
\end{figure}

Fig. \ref{running_time} (a) and (b) draw the training and testing time averaged from 5 runs for each framework. The time usage is approximately proportional to the budget size. DDQN based models even maintain much lower time usage (both training and testing) compared to DQN based models, which  demonstrates the eff\mbox{i}ciency of DDQN based models.

\section{Conclusion}\label{Conclusion}
In this paper, we present a novel end-to-end framework, ToupleGDD, to address the IM problem by leveraging DRL technique. Specif\mbox{i}cally, we incorporates graph neural networks for network embedding and RL technique, double deep Q-networks, for parameters learning. Compared to the state-of-the-art sampling-based approximation algorithms, ToupleGDD can avoid costly sampling of the diffusion paths. Compared to previous works using DRL method for the IM problem, our model have a stronger generalization ability and show almost consistent performance across different social networks. We conduct extensive experiments to evaluate the performance of our proposed model. The empirical results show that ToupleGDD can achieve almost equal expected spread to that of IMM and outperform OPIM-C on several datasets, which is much better than other learning based methods. This validates the effectiveness and eff\mbox{i}ciency of the proposed ToupleGDD model.

\section*{Acknowledgment}
This work was supported in part by NSF under Grant No. 1907472 and No. 1822985, and National Natural Science Foundation of China (NSFC) under Grant No. 62202055.

\bibliographystyle{IEEEtran}
\bibliography{IEEEabrv,reference}

\vspace{-0.5cm}

\begin{IEEEbiography}[{\includegraphics[width=1in,height=1.25in,clip,keepaspectratio]{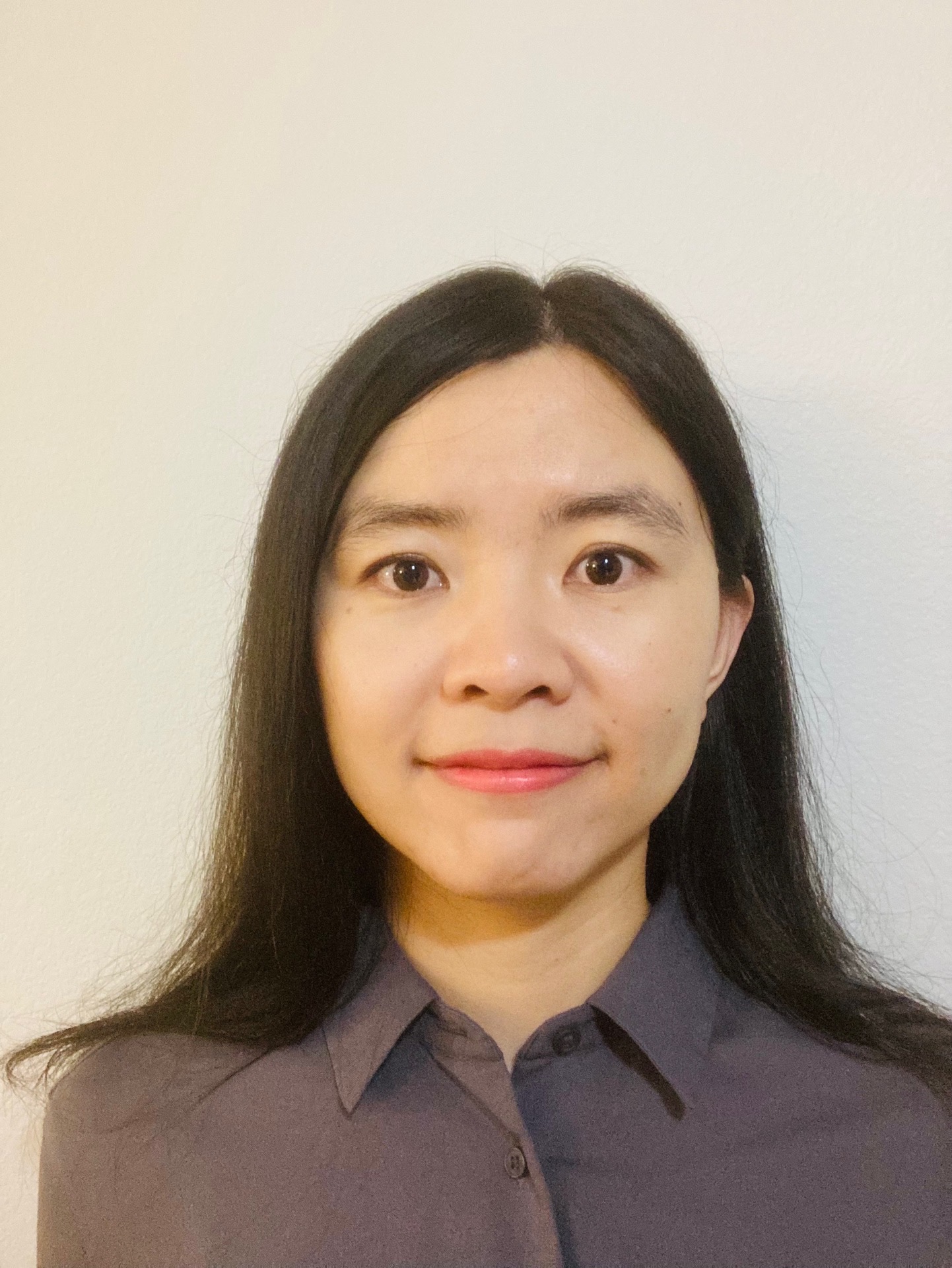}}]{Tiantian Chen}{\space}is a Ph.D. candidate in the Department of Computer Science, The University of Texas at Dallas. She received her B.S. degree in Mathematics and Applied Mathematics, and M.S. degree in Operational Research and Cybernetics from Ocean University of China in 2016 and 2019, respectively. Her research focuses on reinforcement learning, deep learning, social networks, blockchain, and design and analysis of approximation algorithms.
\end{IEEEbiography}

\vspace{-0.5cm}

\begin{IEEEbiography}[{\includegraphics[width=1in,height=1.25in,clip,keepaspectratio]{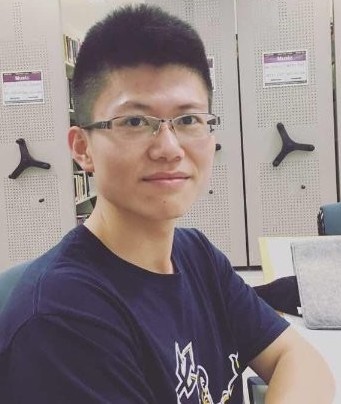}}]{Siwen Yan}{\space}is a Ph.D. candidate in the Department of Computer Science, The University of Texas at Dallas. He received his B.E. degree in Measurement, Control Technique and Instruments from Harbin Institute of Technology in 2015, and M.S. degree in Electrical and Computer Engineering from University of California San Diego in 2017. His research interests include probabilistic graphical models, statistical relational AI, graph neural networks, AI applications in healthcare, reinforcement learning.
\end{IEEEbiography}

\vspace{-0.5cm}

\begin{IEEEbiography}[{\includegraphics[width=1in,height=1.2in,clip,keepaspectratio]{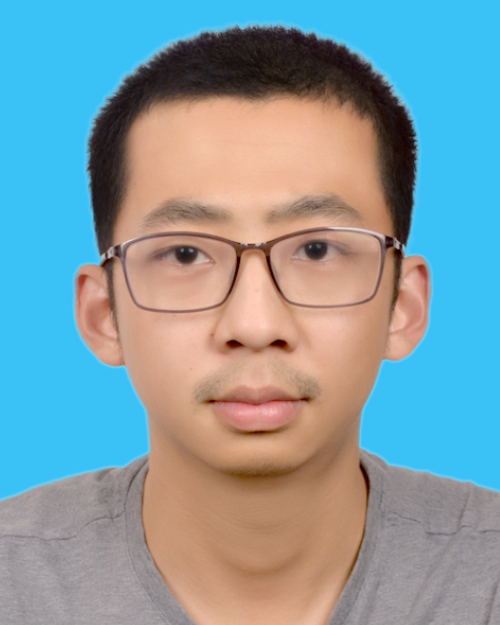}}]{Jianxiong Guo}
received his Ph.D. degree in Computer Science from University of Texas at Dallas in 2021. He is currently an Assistant Professor with the Advanced Institute of Natural Sciences, Beijing Normal University, and also with the Guangdong Key Lab of AI and Multi-Modal Data Processing, BNU-HKBU United International College, Zhuhai, China. He is a member of IEEE/ACM/CCF. He has published more than 40 peer-reviewed papers and been the reviewer for many famous international journals/conferences. His research interests include social networks, wireless sensor networks, combinatorial optimization, and machine learning.
\end{IEEEbiography}

\vspace{-0.5cm}

\begin{IEEEbiography}[{\includegraphics[width=1in,height=1.25in,clip,keepaspectratio]{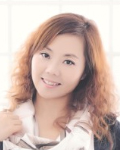}}]{Weili Wu}
 received the Ph.D. and M.S. degrees from the Department of Computer Science, University of Minnesota, Minneapolis, MN, USA, in 2002 and 1998, respectively. She is currently a Full Professor with the Department of Computer Science, The University of Texas at Dallas. Her research mainly deals in the general research area of data communication and data management. Her research focuses on the design and analysis of algorithms for optimization problems that occur in wireless networking environments and various database systems.
\end{IEEEbiography}

\end{document}